\documentstyle[11pt,epsfig]{article}
\begin{document}
\title{\Large\bf The newly observed open-charm states in quark model}

\author{\small De-Min Li\footnote{lidm@zzu.edu.cn}, Peng-Fei Ji, and Bing Ma, \\
\small Department of Physics, Zhengzhou University, Zhengzhou, Henan
450052, People's Republic of China }

\date{\today}

\maketitle

\vspace{0.5cm}

\begin{abstract}

Comparing the measured properties of the newly observed open-charm
states $D(2550)$, $D(2600)$, $D(2750)$, $D(2760)$, $D_{s1}(2710)$,
$D_{sJ}(2860)$, and $D_{sJ}(3040)$ with our predicted spectroscopy
and strong decays in a constituent quark model, we find that: (1)
the $D(2\,^1S_0)$ assignment to $D(2550)$ remains open for its too
broad width determined by experiment; (2) the $D(2600)$ and
$D_{s1}(2710)$ can be identified as the $2\,^3S_1$-$1\,^3D_1$
mixtures; (3) if the $D(2760)$ and $D(2750)$ are indeed the same
resonance, they would be the $D(1\,^3D_3)$; otherwise, they could be
assigned as the $D(1\,^3D_3)$ and $D^\prime_2(1D)$, respectively;
(4) the $D_{sJ}(2860)$ could be either the $D_{s1}(2710)$'s partner
or the $D_s(1\,^3D_3)$; and (5) both the $D_{s1}(2P)$ and
$D^\prime_{s1}(2P)$ interpretations for the $D_{sJ}(3040)$ seem
likely. The $E1$ and $M1$ radiative decays of these sates are also
studied. Further experimental efforts are needed to test the present
quarkonium assignments for these new open-charm states.

\end{abstract}

\vspace{0.5cm}

 {\bf PACS numbers:} 12.39.-x, 13.20.Fc, 13.25.Ft, 14.40.Lb

\newpage

\baselineskip 24pt

\section*{I. Introduction}
\indent \vspace*{-1cm}

In 2009, in inclusive $e^+e^-$ interactions, two new charmed-strange
states $D_{s1}(2710)$ and $D_{sJ}(2860)$ were observed by the BaBar
Collaboration in both $DK$ and $D^\ast K$ channels\cite{babarcs}.
The available experimental results on these two states are as
follows:
\begin{eqnarray}
&&M(D_{sJ}(2860)^+)=2862\pm 2^{+5}_{-2}~\mbox{MeV},
\Gamma(D_{sJ}(2860)^+)=48\pm 3\pm 6 ~\mbox{MeV},\\
&&M(D_{s1}(2710)^+)=2710\pm 2^{+12}_{-7}~\mbox{MeV},
\Gamma(D_{s1}(2710)^+)=149\pm 7^{+39}_{-52}~ \mbox{MeV},\\
&&\frac{{\cal{B}}(D_{s1}(2710)^+\rightarrow D^\ast
K)}{{\cal{B}}(D_{s1}(2710)^+\rightarrow D K)}=0.91\pm
0.13\pm 0.12,\\
&&\frac{{\cal{B}}(D_{sJ}(2860)^+\rightarrow D^\ast
K)}{{\cal{B}}(D_{sJ}(2860)^+\rightarrow D K)}=1.10\pm 0.15\pm 0.19,
\end{eqnarray}
with $DK=D^+K^0+D^0K^+$ and $D^\ast K={D^\ast}^+K^0+{D^\ast}^0K^+$.
In $D^\ast K$ channel, the BaBar Collaboration also found the
evidence for the $D_{sJ}(3040)$ whose mass and width are $3044\pm 8
^{+30}_{-5}$ MeV and $239\pm 35 ^{+46}_{-42}$ MeV, respectively.
There is no signal of $D_{sJ}(3040)$ in $DK$ channel\cite{babarcs}.

More recently, in inclusive $e^+e^-$ collisions, four new charmed
states $D(2550)$, $D(2600)$, $D(2750)$, and $D(2760)$ were found by
the BaBar Collaboration\cite{babarcn}. The $D(2550)$ and $D(2750)$
were observed in ${D^\ast}^+\pi^-$ channel, the $D(2760)$ was
observed in $D^+\pi^-$ channel, and the $D(2600)$ was observed in
both ${D^\ast}^+\pi^-$ and $D^+\pi^-$ channels. The isospin partners
of the $D(2600)^0$ and $D(2760)^0$ were also observed in the
$D^0\pi^+$ channel. The resulting masses and widths of these four
states with neutral-charge are
\begin{eqnarray}
&&M(D(2550)^0)=2539.4\pm 4.5\pm 6.8~\mbox{MeV},
\Gamma(D(2550)^0)=130\pm 12\pm 13 ~\mbox{MeV},\\
&&M(D(2600)^0)=2608.7\pm 2.4\pm 2.5~\mbox{MeV},
\Gamma(D(2600)^0)=93\pm 6\pm 13~ \mbox{MeV},\\
&&M(D(2760)^0)=2763.3\pm 2.3\pm 2.3~\mbox{MeV},
\Gamma(D(2760)^0)=60.9\pm 5.1\pm 3.6~ \mbox{MeV},\\
&&M(D(2750)^0)=2752.4\pm 1.7\pm 2.7~\mbox{MeV},
\Gamma(D(2750)^0)=71\pm 6\pm 11 ~\mbox{MeV},
\end{eqnarray}
and the following ratios of branching fractions were also obtained :
\begin{eqnarray}
\frac{{\cal{B}}(D(2600)^0\rightarrow D^+
\pi^-)}{{\cal{B}}(D(2600)^0\rightarrow D^{\ast +} \pi^-)}=0.32\pm
0.02\pm 0.09,\\
\frac{{\cal{B}}(D(2760)^0\rightarrow D^+
\pi^-)}{{\cal{B}}(D(2750)^0\rightarrow D^{\ast +} \pi^-)}=0.42\pm
0.05\pm 0.11.
\end{eqnarray}

Due to the poor information on the higher excitations of  $D$ and
$D_s$ mesons, the find of these open-charm states is clearly of
importance to complete the $D$ and $D_s$ spectra.  To understand
their observed properties, various efforts have been carried out
under the assumption that all the observed open-charm states are
dominated by the simple $q\bar{q}$ quark
content\cite{close06,mat07,li07,zhang09,close06,liu07x,ruppx,colx,
colx1,zhong08,sun09,zhong09,li09,sun10,zhong10,wang10}. It is
natural and necessary to exhaust the possible conventional
$q\bar{q}$ descriptions before resorting to more exotic
interpretations\cite{exotic}. Further theoretical efforts are still
required in order to satisfactorily explain the data concerning
these open-charm states. In this work, we shall investigate the
masses as well as strong and radiative decays of these newly
observed states in the nonrelativistic constituent quark model and
try to clarify their possible quarkonium assignments by comparing
our predictions with the experiment.

The organization of this paper is as follows. In Sec. II, we
calculate the open-charm mesons masses in a nonrelativistic
constituent quark model and give the possible assignments for these
open-charm states based on their observed masses and decay modes. In
Sec. III, we investigate, with the $^3P_0$ decay model, the strong
decays of these states for different possible assignments. The
radiative transitions of these states are given in Sec. IV. The
summary and conclusion are given in Sec. V.

\section*{II. Masses }
\indent\vspace*{-1cm}

To estimate the masses of $c\bar{u}$ and $c\bar{s}$ states, we
employ a simple nonrelativistic constituent quark model which was
proposed by Lakhina and Swanson and turns out to be able to describe
the heavy-light meson and the charmonium masses with reasonable
accuracy.\cite{swansonmodel}. In this model, the Hamiltonian  is
\begin{eqnarray}
H=H_0+H_{sd}+C_{q\bar{q}}, \label{cqm}
\end{eqnarray}
where $H_0$ is the zeroth-order Hamiltonian, $H_{sd}$ is the
spin-dependent Hamiltonian, and $C_{q\bar{q}}$ is a constant. The
$H_0$ is
\begin{eqnarray}
H_0=\frac{\bf
{P}^2}{M_r}-\frac{4}{3}\frac{\alpha_s}{r}+br+\frac{32\alpha_s\sigma^3e^{-\sigma^2r^2}}{9\sqrt{\pi}m_qm_{\bar{q}}}
{\bf S}_q\cdot{{\bf S}_{\bar{q}}},
\end{eqnarray}
where $r$=$|{\bf r}|$ is the $q\bar{q}$ separation, $M_r=
2m_qm_{\bar{q}}/(m_q+m_{\bar{q}})$; $m_q$ and ${\bf S}_{q}$
($m_{\bar{q}}$ and ${\bf S}_{\bar{q}}$) are the mass and spin of the
constituent quark $q$ (antiquark ${\bar{q}}$), respectively. The
$H_{sd}$ is
\begin{eqnarray}
H_{sd}&=&\left(\frac{{\bf S}_q}{2m^2_q}+\frac{{\bf S}_{
\bar{q}}}{2m^2_{\bar{q}}}\right)\cdot{{\bf
L}}\left(\frac{1}{r}\frac{dV_c}{dr}+\frac{2}{r}\frac{dV_1}{dr}\right)+
\frac{{\bf S_+}\cdot{{\bf L}}}{m_qm_{\bar{q}}} \left(\frac{1}{r}\frac{dV_2}{r}\right)\nonumber\\
&&+\frac{3{\bf S}_q\cdot{\hat{{\bf r}}}{\bf S}_{\bar{q}}\cdot{\hat{{
\bf r}}}-{\bf S}_q\cdot{{\bf
S}_{\bar{q}}}}{3m_qm_{\bar{q}}}V_3+\left [\left(\frac{{\bf
S}_q}{m^2_q}-\frac{{\bf
S}_{\bar{q}}}{m^2_{\bar{q}}}\right)+\frac{{\bf S}_-}{m_qm_{\bar{q}}}
\right ]\cdot{{\bf L}}V_4.
\end{eqnarray}
Here ${\bf L}$ is the relative orbital angular momentum between $q$
and $\bar{q}$, and
\begin{eqnarray}
&&V_c=-\frac{4}{3}\frac{\alpha_s}{r}+br,\nonumber\\
&&V_1=-br-\frac{2}{9\pi}\frac{\alpha^2_s}{r}\left
[9\ln(\sqrt{m_qm_{\bar{q}}}~r)+9\gamma_E-4\right
],\nonumber\\
&&V_2=-\frac{4}{3}\frac{\alpha_s}{r}-\frac{1}{9\pi}\frac{\alpha^2_s}{r}\left
[ -18\ln(\sqrt{m_qm_{\bar{q}}}~r)+54\ln(\mu r)+36\gamma_E+29\right
],\nonumber\\
&&V_3=\frac{4\alpha_s}{r^3}+\frac{1}{3\pi}\frac{\alpha^2_s}{r^3}\left
[-36\ln(\sqrt{m_qm_{\bar{q}}}~r)+54\ln(\mu r)+18\gamma_E+31\right
],\nonumber\\
&&V_4=\frac{1}{\pi}\frac{\alpha^2_s}{r^3}\ln\left(\frac{m_{\bar{q}}}{m_q}\right),\nonumber\\
&&{\bf S}_{\pm}={\bf S}_q\pm{\bf S}_{\bar{q}},
\end{eqnarray}
where $\gamma_E=0.5772$ and the scale $\mu$ has been set to 1.3 GeV.

 The model parameters have been chosen to reproduce the low lying $D$ and $D_s$ masses and are
 $\alpha_s=0.5$, $b=0.14$ GeV$^2$,
$\sigma=1.17$ GeV,
 $C_{c\bar{u}}=-0.325$ GeV, and $C_{c\bar{s}}=-0.275$ GeV. The
constituent quark masses are taken to be
 $m_c=1.43$ GeV,
 $m_u=m_d=0.45$ GeV, and $m_s=0.55$ GeV. These quark
 masses are also used in both strong and radiative decays
 computations.

The heavy-light mesons are not the charge conjugation eigenstates
and hence mixing can occur between the two states with $J=L$. This
mixing can be parameterized as\cite{gi}
\begin{eqnarray}
\left(\begin{array}{c}
c\bar{q}~(nL)\\\
c\bar{q}~^\prime(nL)\end{array}\right) =\left(\begin{array}{cc}
\cos\phi^{c\bar{q}}_L&\sin\phi^{c\bar{q}}_L\\-\sin\phi^{c\bar{q}}_L&\cos\phi^{c\bar{q}}_L\end{array}\right)\left(\begin{array}{c}
n\,^1L_L\\n\,^3L_L\end{array}\right),
\end{eqnarray}
where $\phi$ is the mixing angle and $q$ denotes $u$ or $s$ quark.
The $c\bar{q}~^\prime(nL)$ refers to the higher mass state.

 With the help of the Mathematica program\cite{mathp}, solving the
 Schr$\ddot{\mbox{o}}$dinger equation with Hamiltonian $H_0$ and
 evaluating the $H_{sd}$ in leading-order perturbation theory, one
 can obtain the open charm mesons masses as shown in Tables
 \ref{tab:mesonmass}-\ref{tab:mesonmass1}.\footnote{The mixing angles in radians are
 $\phi^{c\bar{u}}_{1P}=0.363$,
 $\phi^{c\bar{s}}_{1P}=0.427$,
 $\phi^{c\bar{u}}_{2P}=0.578$,
 $\phi^{c\bar{s}}_{2P}=0.564$,
 $\phi^{c\bar{u}}_{1D}=0.697$,
 $\phi^{c\bar{s}}_{1D}=0.701$,
 $\phi^{c\bar{u}}_{2D}=0.702$,
   and $\phi^{c\bar{s}}_{2D}=0.708$.}
 For comparison, the corresponding masses predicted by some other approaches such as
 the Blankenbecler-Sugar equation\cite{lnr} and the relativistic quark
 model\cite{gi,zvr,efg,de} are also listed.

{\small
\begin{table}[hbt]
\begin{center}
\vspace*{-0.5cm}
 \caption{\label{tab:mesonmass}\small The charmed meson masses in GeV.}

 \vspace*{0.2cm}
\begin{tabular}{lcccccccc}\hline\hline
  State  &             $J^P$     &$\mbox{This work}$    &LNR\cite{lnr}&ZVR\cite{zvr}&EFG\cite{efg}&DE\cite{de}&GI\cite{gi}       &PDG\cite{pdg08} \\ \hline
  $D(1\,^1S_0)$        &$0^-$       &1.867               &1.874       &1.85         &1.871        &1.868      &1.88               &1.869      \\
  $D(2\,^1S_0)$        &$0^-$       &2.555               &2.540       &2.50         &2.581        &2.589      &2.58               &       \\

  $D(1\,^3S_1)$        &$1^-$       &2.010               &2.006       &2.02         &2.010        &2.005      &2.04               &2.010      \\
  $D(2\,^3S_1)$        &$1^-$       &2.636               &2.601       &2.62         &2.632        &2.692      &2.64               &       \\

  $D(1\,^3P_0)$        &$0^+$       &2.252               &2.341       &2.27         &2.406        &2.377      &2.40               &2.308      \\
  $D(2\,^3P_0)$        &$0^+$       &2.752               &2.758       &2.78         &2.919        &2.949      &                   &       \\

  $D(1\,^3P_2)$        &$2^+$       &2.466               &2.477       &2.46         &2.460        &2.460      &2.50               &2.460      \\
  $D(2\,^3P_2)$        &$2^+$       &2.971               &2.860       &2.94         &3.012        &3.035      &                  &       \\

  $D_1(1P)$            &$1^+$       &2.402               &2.389       &2.40         &2.426        &2.417      &2.44               &2.427      \\
  $D^\prime_1(1P)$     &$1^+$       &2.417               &2.407       &2.41         &2.469        &2.490      &2.49               &2.422      \\

  $D_1(2P)$            &$1^+$       &2.886               &2.792       &2.89         &2.932        &2.995      &                &       \\
  $D^\prime_1(2P)$     &$1^+$       &2.926               &2.802       &2.90         &3.021        &3.045      &                &       \\

  $D(1\,^3D_1)$        &$1^-$       &2.740               &2.750       &2.71         &2.788        &2.795      &2.82                &           \\
  $D(2\,^3D_1)$        &$1^-$       &3.168               &3.052       &3.13         &3.228        &           &                 &           \\

  $D(1\,^3D_3)$        &$3^-$       &2.719               &2.688       &2.78         &2.863        &2.799      &2.83                &          \\
  $D(2\,^3D_3)$        &$3^-$       &3.170               &2.999       &3.19         &3.335        &           &                 &          \\

  $D_2(1D)$            &$2^-$       &2.693               &2.689       &2.74         &2.806        &2.775      &                 &           \\
  $D^\prime_2(1D)$     &$2^-$       &2.789               &2.727       &2.76         &2.850        &2.833      &                 &           \\

  $D_2(2D)$            &$2^-$       &3.145               &2.997       &3.16         &3.259        &           &                &           \\
  $D^\prime_2(2D)$     &$2^-$       &3.215               &3.029       &3.17         &3.307        &           &                &           \\
\hline\hline
\end{tabular}
\end{center}
\end{table}
}

{\small
\begin{table}[hbt]
\begin{center}
\vspace*{-0.5cm}
 \caption{\label{tab:mesonmass1}\small The charm-strange meson masses in GeV.}
 \vspace*{0.2cm}
\begin{tabular}{lcccccccc}\hline\hline
  State  &             $J^P$     &$\mbox{This work}$      &LNR\cite{lnr}&ZVR\cite{zvr}  &EFG\cite{efg}&DE\cite{de}  &GI\cite{gi}     &PDG\cite{pdg08} \\ \hline
  $D_s(1\,^1S_0)$        &$0^-$       &1.969              &1.975          &1.94         &1.969        &1.965        &1.98           &1.969      \\
  $D_s(2\,^1S_0)$        &$0^-$       &2.640              &2.659          &2.61         &2.688        &2.700        &2.67           &       \\

  $D_s(1\,^3S_1)$        &$1^-$       &2.107              &2.108          &2.13         &2.111        &2.113        &2.13           &2.112      \\
  $D_s(2\,^3S_1)$        &$1^-$       &2.714              &2.722          &2.73         &2.731        &2.806        &2.73           &       \\

  $D_s(1\,^3P_0)$        &$0^+$       &2.344              &2.455          &2.38         &2.509        &2.487        &2.48           &2.317      \\
  $D_s(2\,^3P_0)$        &$0^+$       &2.830              &2.901          &2.90         &3.054        &3.067        &               &       \\

  $D_s(1\,^3P_2)$        &$2^+$       &2.559              &2.586          &2.58         &2.571        &2.581        &2.59            &2.572      \\
  $D_s(2\,^3P_2)$        &$2^+$       &3.040              &2.988          &3.06         &3.142        &3.157        &               &       \\

  $D_{s1}(1P)$           &$1^+$       &2.488              &2.502          &2.51         &2.536        &2.535        &2.53           &2.459      \\
  $D^\prime_{s1}(1P)$    &$1^+$       &2.510              &2.522          &2.52         &2.574        &2.605        &2.57           &2.535      \\

  $D_{s1}(2P)$           &$1^+$       &2.958              &2.928          &3.00         &3.067        &3.114        &               &       \\
  $D^\prime_{s1}(2P)$    &$1^+$       &2.995              &2.942          &3.01         &3.154        &3.165        &               &       \\

  $D_s(1\,^3D_1)$        &$1^-$       &2.804              &2.845          &2.82         &2.913        &2.913        &2.90           &           \\
  $D_s(2\,^3D_1)$        &$1^-$       &3.217              &3.172          &3.25         &3.383        &             &               &           \\

  $D_s(1\,^3D_3)$        &$3^-$       &2.811              &2.844          &2.90         &2.917        &2.925        &2.92            &          \\
  $D_s(2\,^3D_3)$        &$3^-$       &3.240              &3.157          &3.31         &3.469        &             &                &          \\

  $D_{s2}(1D)$           &$2^-$       &2.788              &2.817          &2.86         &2.931        &2.900        &                &           \\
  $D^\prime_{s2}(1D)$    &$2^-$       &2.849              &2.844          &2.88         &2.961        &2.953        &                &           \\

  $D_{s2}(2D)$           &$2^-$       &3.217              &3.144          &3.28         &3.403        &             &               &           \\
  $D^\prime_{s2}(2D)$    &$2^-$       &3.260              &3.167          &3.29         &3.456        &             &               &           \\
\hline\hline
\end{tabular}
\end{center}
\end{table}
} {\small
\begin{table}[hbt]
\begin{center}
\vspace*{-0.5cm}
 \caption{\label{tab:jp}\small Possible $J^P$ of the open-charm states based on the observed decay modes.}
 \vspace*{0.2cm}
\begin{tabular}{lcc}\hline\hline
  State              & observed channel   &Possible $J^P$\\\hline
  $D(2550)$          &$D^\ast\pi$         &$0^-$, $1^-$, $2^-$$\cdots$\\
  $D(2600)$          &$D\pi$, $D^\ast\pi$ &$1^-$, $3^-$$\cdots$\\
  $D(2750)$          &$D^\ast\pi$         &$0^-$, $1^-$, $2^-$, $3^-$$\cdots$\\
  $D(2760)$          &$D\pi$              &$0^+$, $1^-$, $2^+$, $3^-$$\cdots$\\
  $D_{sJ}(2860)$     &$DK$, $D^\ast K$    &$1^-$, $3^-$$\cdots$\\
  $D_{sJ}(3040)$     &$D^\ast K$          &$0^-$, $1^+$, $2^-$$\cdots$\\
  \hline\hline
  \end{tabular}
  \end{center}
  \end{table}
}

 It is clear from Tables
\ref{tab:mesonmass} and \ref{tab:mesonmass1} that the quark model
(\ref{cqm}) can reasonably account for the masses of the observed
ground $S$ and $P$-wave open-charm mesons, and the overall agreement
between the expectations from the quark model (\ref{cqm}) and those
from other approaches, especially the Blankenbecler-Sugar
equation\cite{lnr} and the relativistic quark model\cite{zvr}, is
good, which hence encourages us to discuss the possible assignments
for the newly observed open-charm states based on the expectations
of our employed quark model. Among these newly observed open-charm
states, the $J^P$ of $D_{s1}(2710)$ is determined to be $1^-$
experimentally\cite{belle1}, while the spin-parity quantum numbers
of the other states are still unsettled. According to the observed
decay modes, the possible spin-parity quantum numbers of these
open-charm states are listed in Table \ref{tab:jp}.

We shall discuss the possible quarkonium assignments for these
open-charm states based on Tables \ref{tab:mesonmass},
\ref{tab:mesonmass1} and \ref{tab:jp}. The possible $J^P$ of
$D(2550)$ are $0^-$, $1^-$, $2-$, $\cdots$. The
$1^-[D(2\,^3S_1,1\,^3D_1)]$ and $2^-[D^{(\prime)}_{s2}(2D)]$ are
expected to be at least about 100 MeV higher than $D(2550)$ in
mass. the $1^-$ and $2^-$ assignments to $D(2550)$ are
implausible. The $D(2550)$ mass is very close to the predicted
mass for the $0^-$[$D(2\,^1S_0)$](2555 MeV) and the helicity-angle
distribution of $D(2550)$ turns out to be consistent with the
predictions for $D(2\,^1S_0)$\cite{babarcn}. Therefore, the
$0^-[D(2\,^1S_0)]$ assignment to $D(2550)$ seems the most
plausible. The possible $J^P$ of $D(2600)$ are $1^-$, $3^-$,
$\cdots$. The $D(2600)$ mass is very close to the predicted mass
for the $1^-[D(2\,^3S_1)]$ (2636 MeV). Also, the $D(2\,^3S_1$) and
$D(1\,^3D_1)$ have the same $J^P$ and similar masses, and hence in
general cam mix to produce two physical $1^-$
states\footnote{Hereafter, we shall assign the $1^-$ physical
states as the $2\,^3S_1$-$1\,^3D_1$ mixtures.}. Therefore, the
$D(2600)$ is most likely the $2\,^3S_1$-$1\,^3D_1$ mixtures
$1^-[D$(2S-1D)]. The helicity-angle distribution of $D(2600)$ is
found to be also consistent with the predicted $D(2\,^3S_1)$ or
$D(1\,^3D_1)$\cite{babarcn}. Similarly, the $D_{s1}(2710)$ is most
likely the $1^-[D_s$(2S-1D)]. The possible $J^P$ of $D(2750)$
include $0^-$, $1^-$, $2^-$, $3^-$, $\cdots$. The $D(2750)$ mass
is about 200 MeV higher than that of $0^-[D(2\,^1S_0)]$, which
makes the $0^-$ assignment to $D(2750)$ implausible. The $D(2750)$
mass is very close to the masses of $1^-[D(1\,^3D_1)]$(2740 MeV),
$3^-[D(1\,^3D_3)]$(2719 MeV), $2^-[D_2(1D)]$ (2693 MeV), and
$2^-[D^\prime_2(1D)]$ (2789 MeV). Therefore, the plausible
assignments for $D(2750)$ would be $1^-[D$(2S-1D)],
$3^-[D(1\,^3D_3)]$, and $2^-[D_2(1D), ~D^\prime_2(1D)]$.  The
possible $J^P$ of $D(2760)$ include $0^+$, $1^-$, $2^+$, $3^-$,
$\cdots$. The mass difference between the $D(2760)$ and
$2^+[D(1\,^3P_2,2\,^3P_2)]$ is about 200 MeV while the $D(2750)$
mass is very close to the masses of $1^-[D(1\,^3D_1)]$(2740 MeV),
$3^-[D(1\,^3D_3)]$(2719 MeV), and $0^+[D(2\,^3P_0)]$ (2752 MeV),
which makes $1^-[D$(2S-1D)], $3^-D(1\,^3D_3)$, and
$0^+D(2\,^3P_0)]$ assignments to $D(2760)$ possible. The possible
$J^P$ of the $D_{sJ}(2860)$ are $1^-$, $3^-$, $\cdots$. Since the
$D_{sJ}(2860)$ mass is close to $1^-[D_s(1\,^3D_1)]$ mass (2804
MeV) and $3^-[D_s(1\,^3D_3)]$ mass (2811 MeV), the possible
assignments for $D_{sJ}(2860)$ include $1^-[D_s$(2S-1D)] and
$3^-[D_s(1\,^3D_3)]$. The possible $J^P$ of $D_{sJ}(3040)$ are
$0^-$, $1^+$, $2^-$, $\cdots$. The $D_{sJ}(3040)$ is far higher
than the $0^-[D_s(2\,^3S_1)]$ and $2^-[D^{(\prime)}_{s2}(1D)]$ in
mass. Also, the $2^-[D^{(\prime)}_{s2}(2D)]$ mass is about 200 MeV
higher than that of $D_{sJ}(3040)$. The possibility of
$D_{sJ}(3040)$ being the $0^-$ and $2^-$ can be ruled out. The
$D_{sJ}(3040)$ mass is close to the quark model expectations for
the $1^+[D^{(\prime)}_{s1}(1P)]$ mass, therefore the
$1^+[D^{(\prime)}_{s1}(1P)]$ assignment to $D_{sJ}(3040)$ becomes
the most possible.

 Below, we shall focus on these
possible assignments for the observed open-charm states as shown in
Table \ref{tab:assignment}. The mass information alone is
insufficient to classify these new open-charm states. Their decay
properties also need to be compared with model expectations. We
shall discuss the decay dynamics of these states in next section.

{\small
\begin{table}[hbt]
\begin{center}
\vspace*{-0.5cm}
 \caption{\label{tab:assignment}\small Possible assignments of the open-charm states based on their masses and decay modes.}
 \vspace*{0.2cm}
\begin{tabular}{lc}\hline\hline
  State               &Assignments\\\hline
  $D(2550)$           &$0^-$[$D(2\,^1S_0)$]\\
  $D(2600)$           &$1^-$[$D$(2S-1D)]\\
  $D(2750)$           &$1^-$[$D$(2S-1D)], $2^-$[$D_2(1D)$,~$D^\prime_2(1D)$], $3^-$[$D(1\,^3D_3)$]\\
  $D(2760)$           &$1^-$[$D$(2S-1D)], $3^-$[$D(1\,^3D_3)]$, $0^+$[$D(2\,^3P_0)$]\\
  $D_{s1}(2710)$      &$1^-$[$D_s$(2S-1D)]\\
  $D_{sJ}(2860)$      &$1^-$[$D_s$(2S-1D)], $3^-$[$D_s(1\,^3D_3)$]\\
  $D_{sJ}(3040)$     &$1^+$[$D_{s1}(2P)$,~$D^\prime_{s1}(2P)$]\\

\hline\hline
  \end{tabular}
  \end{center}
  \end{table}
}

\section*{III. Strong decays}
\indent \vspace*{-1cm}

\subsection*{A. Model parameters}

\indent\vspace*{-1cm}

In this section, we shall employ the $^3P_0$ model to evaluate the
tow-body open-flavor strong decays of the initial state. The $^3P_0$
model, also known as the quark pair creation model, has been
extensively applied to evaluate the strong decays of mesons from
light $q\bar{q}$ to heavy $c\bar{b}$, since it gives a considerably
good description of many observed decay amplitudes and partial
widths of hadrons. Some detailed reviews on the $^3P_0$ model can be
found in Refs.\cite{3p0rev1,3p0rev2,3p0rev3,3p0rev4,3p0rev5}. Also,
the simple harmonic oscillator (SHO) approximation for spatial wave
functions of mesons is used in the strong decays computations. This
is typical of strong decay calculations. The SHO wave functions have
the advantage that decay amplitudes and widths can be determined
analytically, and it has been demonstrated that the numerical
results are usually not strongly dependent on the details of the
spatial wave functions of mesons\cite{3p0rev5,sho1,sho2,sho3}. The
explicit expression for the decay width employed in this work can be
seen in Refs.\cite{lix}.

The parameters involved in the $^3P_0$ model include the constituent
quarks masses, the SHO wave function scale parameters $\beta$'s, and
the light nonstrange quark pair creation strength $\gamma$. The
$\gamma$ and the strange quark pair creation strength
$\gamma_{s\bar{s}}$ can be related by
$\gamma_{s\bar{s}}\approx\gamma/\sqrt{3}$\cite{sscreation}. The
constituent quarks masses $m_u$, $m_d$, $m_s$, and $m_c$ are the
same as those used in the constituent quark model (\ref{cqm}). The
SHO wave function scale parameters are taken to be the effective
$\beta$'s obtained by equating the root mean square radius of the
SHO wave function to that obtained from the nonrelativistic quark
model (\ref{cqm}). The meson effective $\beta$'s used in this work
are listed in Table \ref{tab:beta}. The remaining parameter $\gamma$
is an overall factor in the width.  By fitting to 19
well-established experimental decay widths,\footnote{ The decay
modes used in our fit are [1] $\rho\rightarrow\pi\pi$, [2]
$\phi\rightarrow KK$, [3] $K^\ast\rightarrow K\pi$, [4]
$b_1\rightarrow\omega\pi$, [5] $K^\ast_2\rightarrow K\pi$, [6]
$K^\ast_2\rightarrow K^\ast\pi$, [7] $K^\ast_2\rightarrow K\rho$,
[8] $K^\ast_2\rightarrow K\omega$, [9] $\pi_2(1670)\rightarrow
f_2\pi$, [10] $\pi_2(1670)\rightarrow K^\ast K$, [11]
$\rho_3(1680)\rightarrow\pi\pi$, [12]
$\rho_3(1680)\rightarrow\omega\pi$, [13] $\rho_3(1680)\rightarrow
KK$, [14] $K_3(1780)\rightarrow K\pi$, [15] $K_3(1780)\rightarrow
K\rho$, [16] $K_3(1780)\rightarrow K^\ast\pi$, [17]
$D_2(2460)^0\rightarrow D\pi+D^\ast\pi$, [18]
$D_2(2460)^+\rightarrow D\pi+D^\ast\pi$, and [19] $D_{s2}\rightarrow
DK+D^\ast K+D_s\eta$. The corresponding data are from
PDG\cite{pdg08}.} we obtain $\gamma=0.452\pm 0.105$, consistent with
$0.485\pm 0.15$ obtained by Close and Swanson from their
model\cite{closeq}. The $\gamma$ uncertainty means that the
theoretical width has an uncertainty of $\delta\Gamma\simeq
0.47\Gamma$. It is perhaps no surprise that the prediction has a
larger uncertainty due to the larger errors of data as well as the
decay model which is tuned for strong decays of momenta of hundreds
of MeV.

The meson masses used to determine the phase space and final state
momenta in both strong and radiative decays computations
are\cite{babarcs,babarcn,pdg08} $M_{\pi^\pm}=139.57$ MeV,
$M_{\pi^0}=134.98$ MeV, $M_{K^\pm}= 493.677$ MeV, $M_{K^0}= 497.614$
MeV, $M_{\eta}=547.853$ MeV, $M_\rho=775.49$ MeV, $M_\omega=782.65$
MeV, $M_{{K^\ast}^\pm}=891.66$ MeV, $M_{{K^\ast}^0}=896$ MeV,
$M_{D^\pm }=1869.62$ MeV, $M_{D^0 }=1864.84$ MeV,
$M_{{D^\ast}^\pm}=2010.27$ MeV, $M_{{D^\ast}^0}=2006.97$ MeV,
 $M_{D_s}=1968.49$ MeV, $M_{D^\ast_s}=2112.3$ MeV,
$M_{D_1(2430)^0}=M_{D_1(2430)^\pm}=2427$ MeV,
$M_{D_1(2420)^\pm}=2423.4$ MeV, $M_{D_1(2420)^0}=2422.3$ MeV,
$M_{D_0(2400)^0}=2308$ MeV, $M_{D_0(2400)^\pm}=2403$ MeV,
$M_{D_2(2460)^\pm}=2460.1$ MeV, $M_{D_2(2460)^0}=2461.1$ MeV,
$M_{D_{s1}(2460)}=2459.6$ MeV, $M_{D_{s1}(2535)}=2535.35$ MeV,
$M_{D_{s0}(2317)}=2317.8$ MeV, $M_{D_{s2}(2573)}=2572.6$ MeV,
$M_{D(2550)^0}=2539.4$ MeV, $M_{D(2600)^0}=2608.7$ MeV,
$M_{D(2760)^0}=2763.3$ MeV, $M_{D(2750)^0}=2752.4$ MeV,
$M_{D_{s1}(2710)}=2710$ MeV, $M_{D_{sJ}(2860)}=2862$ MeV, and
$M_{D_{sJ}(3040)}=3044$ MeV.
 The meson flavor functions follow the conventions of Ref.\cite{gi},
 for example, $D^0=c\bar{u}$, $D^+=-c\bar{d}$, $D^+_s=-c\bar{s}$, $K^+=-u\bar{s}$,
 $K^-=s\bar{u}$, $K^0=-d\bar{s}$, $\pi^+=-u\bar{d}$, $\pi^0=(u\bar{u}-d\bar{d})/\sqrt{2}$, $\phi=-s\bar{s}$, $\omega=(u\bar{u}+d\bar{d})/\sqrt{2}$, and
  $\eta=(u\bar{u}+d\bar{d})/2-s\bar{s}/\sqrt{2}$. Also, we set
$D_1(1P)=D_1(2430)$, $D^\prime_1(1P)=D_1(2420)$,
$D_{s1}(1P)=D_{s1}(2460)$, and $D^\prime_{s1}(1P)=D_{s1}(2535)$. The
  $n\,^3L_J$-$n\,^1L_J$ mixing angles are taken as those determined in the mass
  estimates.

   {\small
\begin{table}[hbt]
\begin{center}
\vspace*{-0.5cm}
 \caption{\label{tab:beta}\small The meson effective $\beta$ values in MeV.}
 \vspace*{0.2cm}
\begin{tabular}{cccccc}\hline\hline
  $n\,^{2S+1}L_J$  & $u\bar{u}$ &$u\bar{s}$ & $s\bar{s}$ &$c\bar{u}$&$c\bar{s}$\\\hline

  $1\,^1S_0$&470&466&470&453  &484\\
  $2\,^1S_0$&294&301&310&325  &343\\
  $1\,^3S_1$&308&322&338&379  &406\\
  $2\,^3S_1$&258&267&279&306  &324\\

  $1\,^3P_J$&280&290&302&328  &348\\
  $2\,^3P_J$&247&255&265&287  &303\\

  $1\,^1P_1$&284&294&306&332  &352\\
  $2\,^1P_1$&250&259&269&290  &306\\

  $1\,^3D_J$&261&270 &281 &304  &321\\
  $2\,^3D_J$&238&246 &255 &275  &290\\

  $1\,^1D_2$&261&270 &281 &304  &321\\
  $2\,^1D_2$&238&246 &255 &275  &290\\
  \hline\hline
\end{tabular}
\end{center}
\end{table}
}

\subsection*{B. $D(2550)$}

\indent\vspace*{-1cm}

 The decay widths of $D(2550)$ as $D(2\,^1S_0)$
are shown in Table \ref{tab:pw2550}. The predicted total width is
about 45 MeV, about 70 MeV lower than the lower limit of the
measured $\Gamma(D(2550))=130\pm 12\pm 13$ MeV. The recent
calculations in a $^3P_0$ model\cite{sun10} and a chiral quark
model\cite{zhong10} also give a rather narrow width for the
$D(2\,^1S_0)$. The upper limit of the $D(2\,^1S_0)$'s width is
expected to be about 66 MeV, still about 50 MeV lower than the lower
limit of the measurement. This inconsistency between the theoretical
and experimental results could imply that the experimental analysis
has overestimated the width of $D(2550)$ if this state is indeed the
$2\,^1S_0$ charmed meson, as suggested by Ref.\cite{sun10}. Further
confirmation of its resonance parameters is required to confirm the
$D(2\,^1S_0)$ assignment to $D(2550)$. The ratio
$\Gamma(D_0(2400)\pi)/\Gamma(D^\ast\pi)$ is expected to be about
0.22, which is independent of the parameter $\gamma$ and can also
present a consistent check for this assignment. Without additional
information on $D(2550)$, the $D(2\,^1S_0)$ assignment to $D(2550)$
would remain open.

 {\small
\begin{table}[hbt]
\begin{center}
\vspace*{-0.5cm}
 \caption{\label{tab:pw2550}\small Decay widths of $D(2550)$ as $D(2\,^1S_0)$ in MeV.}
 \vspace*{0.2cm}
\begin{tabular}{cccc}\hline\hline
 ${D^\ast}^+\pi^-$
&${D^\ast}^0\pi^0$ &$D_0(2400)^0\pi^0$&$\mbox{Total}$\\\hline
24.86&12.41&8.09&45.35\\
\hline\hline
 \end{tabular}
\end{center}
 \end{table}}

\subsection*{C. $D(2600)$}
 \indent\vspace*{-1cm}

In the 2S-1D mixing scenario, the eigenvectors of $D_1(2600)$ and
its partner $D_1(M_X)$ can be written as
\begin{eqnarray}
&&|D(2600)\rangle=\cos\theta|2\,^3S_1\rangle-\sin\theta|1\,^3D_1\rangle,\\
&&|D(M_X)\rangle=\sin\theta|2\,^3S_1\rangle+\cos\theta|1\,^3D_1\rangle,
\end{eqnarray}
where the $\theta$ is the $D(2\,^3S_1)$-$D(1\,^3D_1)$ mixing angle
and $M_X$ denotes the mass of the physical state $D_1(M_X)$.

\begin{figure}[hbt]
\begin{center}
\epsfig{file=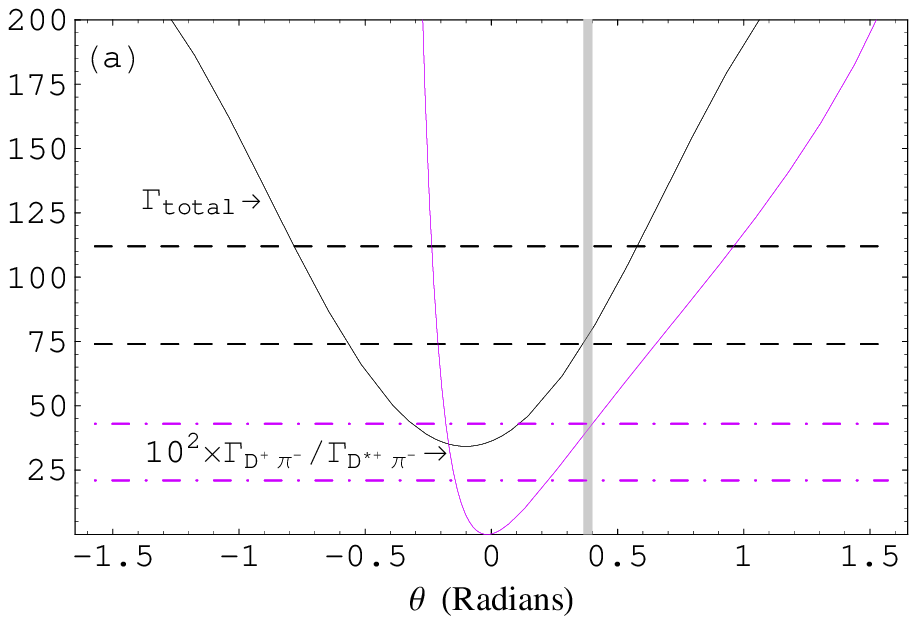,width=6.0cm, clip=}
\epsfig{file=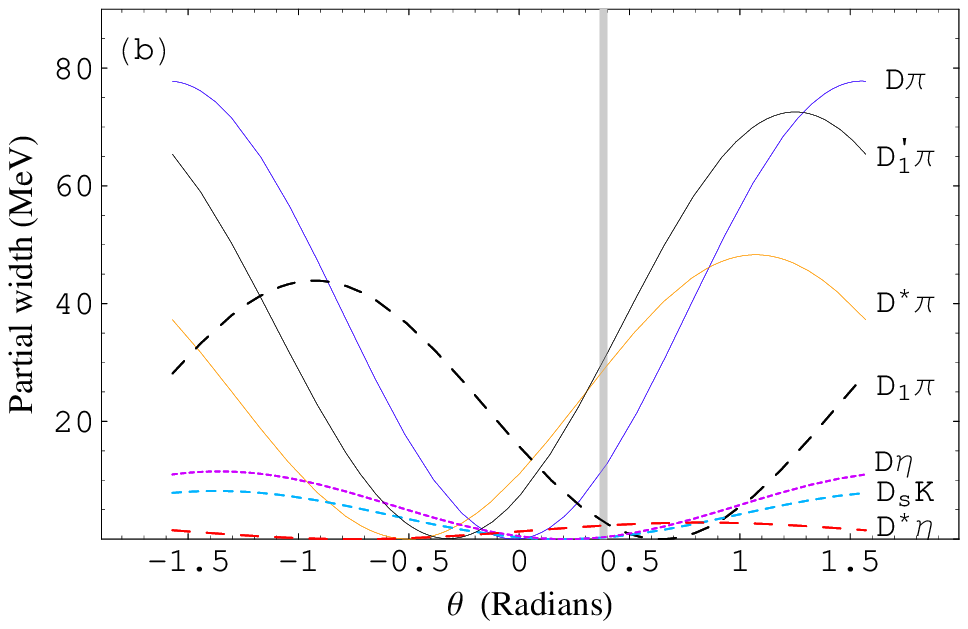,width=6.0cm, clip=}

 \vspace*{-0.3cm}
\caption{\label{fig:pw2600}\small Decay widths and branching ratio
$\Gamma(D^+\pi)/\Gamma({D^\ast}^+\pi^-)$ of $D(2600)$ versus
$\theta$. The horizontal dashed (dot-dashed) lines indicate the
upper and lower values of the experimental data on the total width
(branching ratio). $\Gamma(D^\ast_sK)$ and $\Gamma(D_2(2460)\pi)$
are tiny and not shown.}
\end{center}
\end{figure}

The predicted decay widths of $D(2600)$ are listed in Table
\ref{tab:pw2600}. The variations of decay widths and branching ratio
$\Gamma(D^+\pi)/\Gamma({D^\ast}^+\pi^-)$ with the mixing angle
$\theta$ are illustrated in Fig. \ref{fig:pw2600}. It is clear that
in the presence of about $0.364 \leq \theta \leq 0.4$ radians, both
the total width and branching ratio
$\Gamma(D^+\pi)/\Gamma({D^\ast}^+\pi^-)$ of $D(2600)$ can be well
reproduced (see Fig. \ref{fig:pw2600}(a)). Also in this mixing angle
range, the $D^\ast\pi$, $D_1(2420)\pi$, and $D\pi$ are the dominant
decay modes and the mode ${D^\ast}^+\pi^-$ dominates $D^+\pi^-$ (see
Fig. \ref{fig:pw2600}(b)), consistent with the observation. The
helicity-angle distribution of $D(2600)$ is also found to be
consistent with the predictions for the $D(2\,^3S_1)$ or
$D(1\,^3D_1)$\cite{babarcn}. Therefore the interpretation of
$D(2600)$ as a mixture of the $D(2\,^3S_1)$ and $D(1\,^3D_1)$ seems
convincing. It is expected that
$\Gamma(D_1(2420)\pi)/\Gamma(D^\ast\pi)$ is around 1.0 and
$\Gamma(D_1(2430)\pi)/\Gamma(D^\ast\eta)$ is $1.1\sim 1.4$. Further
experimental study on the $D(2600)$ in the $D_1(2420)\pi$,
$D_1(2430)\pi$, and $D^\ast\eta$ channels can present a consistent
check for this interpretation.

 {\small
\begin{table}[hbt]
\begin{center}
\vspace*{-0.5cm} \caption{\label{tab:pw2600}\small Decay widths of
$D(2600)$ as the $1^-$ state in MeV. $c\equiv\cos\theta$ and
$s\equiv\sin\theta$.}
 \vspace*{0.2cm}
\begin{tabular}{lc}\hline\hline
 Mode         & $\Gamma_i$    \\ \hline
$D^0\pi^0$        & $0.02c^2+1.28cs+25.90s^2$    \\
$D^+\pi^-$        & $0.02c^2+1.77cs+51.89s^2$    \\

$D_sK$            & $0.32c^2-3.17cs+7.86s^2$      \\

$D\eta$           & $0.50c^2-4.70cs+10.99s^2$      \\

${D^\ast}^0\pi^0$   & $3.62c^2+13.42cs+12.45s^2$   \\

${D^\ast}^+\pi^-$   & $7.44c^2+27.16cs+24.79s^2$   \\

${D^\ast}\eta$      & $1.30c^2+2.80cs+1.51s^2$      \\

${D^\ast_s} K$        & $0.01c^2+0.03cs+0.02s^2$       \\

${D_1(2430)}^0\pi^0$  & $5.38c^2-14.36cs+9.58s^2$       \\

${D_1(2430)}^+\pi^-$  & $10.39c^2-27.76cs+18.55s^2$       \\

${D_1(2420)}^0\pi^0$  & $2.51c^2+14.87cs+22.41s^2$      \\

${D_1(2420)}^+\pi^-$  & $4.78c^2+28.47cs+42.96s^2$      \\

${D_2(2460)}^0\pi^0$  & $(0.79c^2+0.98cs+0.31s^2)\times 10^{-3}$       \\
${D_2(2460)}^+\pi^-$  & $(0.73c^2+0.90cs+0.28s^2)\times 10^{-3}$       \\
              &$\Gamma_{\mbox{t}}=36.28c^2+39.82cs+228.90s^2$\\\hline\hline
\end{tabular}
\end{center}
\end{table}
}

\begin{figure}[hbt]
\begin{center}
\epsfig{file=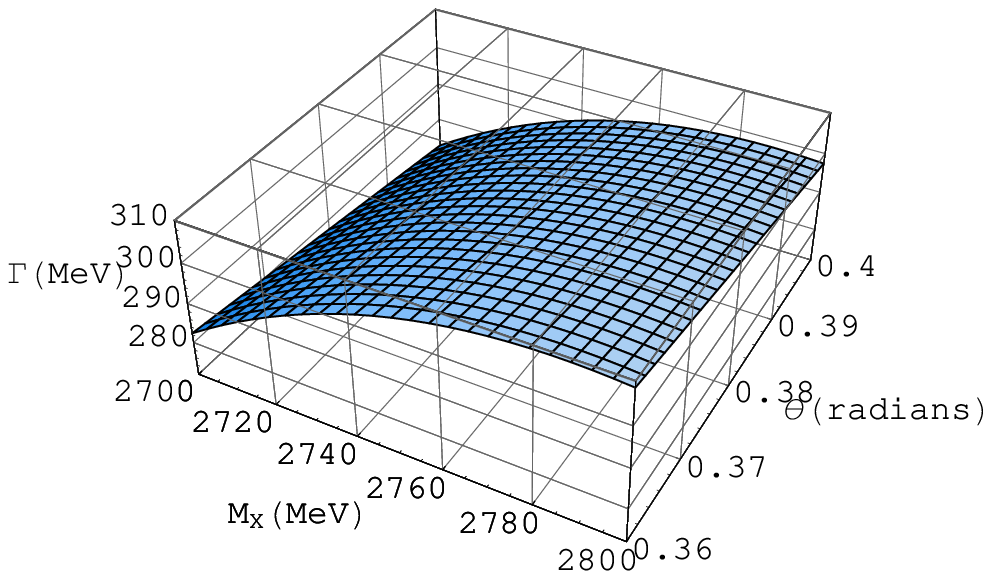,width=6.0cm, clip=}
\epsfig{file=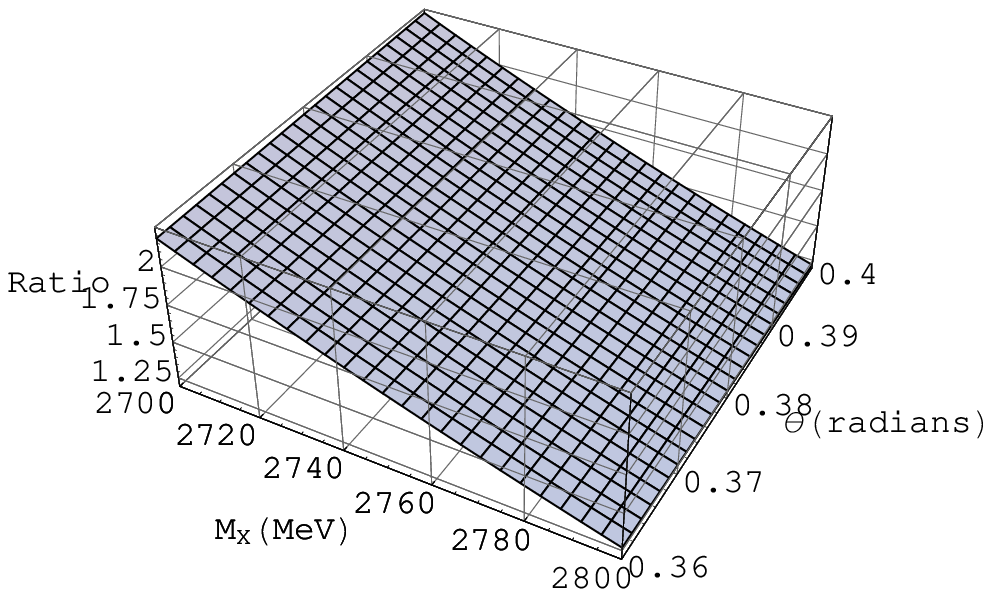,width=6.0cm, clip=}

 \vspace*{-0.3cm}
\caption{\label{fig:predw}\small Total width and branching ratio
$\Gamma(D^+\pi)/\Gamma({D^\ast}^+\pi^-)$ of $D(M_X)$ versus $M_X$
and $\theta$.}
\end{center}
\end{figure}

The $D(M_X)$ is expected to have a mass of about 2.77 GeV according
to $M^2_X= M^2_{D(2\,^3S_1)}+M^2_{D(1\,^3D_1)}-M^2_{D(2600)}$. Other
approaches predicted that the $D(M_X)$ would lie in about
$2.7\sim2.8$ GeV (see Table \ref{tab:mesonmass}). The total width
and branching ratio $\Gamma(D^+\pi^-) /\Gamma({D^\ast}^+\pi^-)$ of
$D(M_X)$ as functions of the initial state mass $M_X$ and the mixing
angle $\theta$ are illustrated in Fig. \ref{fig:predw}. The $M_X$ is
restricted to be $2700\sim 2800$ MeV and $\theta$ is restricted to
be $0.364\sim 0.4$ radians.

As can be seen in Fig. \ref{fig:predw}, with the variations of the
initial state mass and the mixing angle, the total width of $D(M_X)$
varies from about 280 to 310 MeV and the branching ratio
$\Gamma(D^+\pi)/\Gamma({D^\ast}^+\pi^-)$ varies from about 1.25 to
2.25. At around 2760 MeV, the lower limit of the $D(M_X)$'s total
width is expected to be about $140\sim 144$ MeV. Clearly, both the
predicted total width and $\Gamma(D^+\pi)/\Gamma({D^\ast}^+\pi^-)$
of the $D(M_X)$ are in disagreement with the data concerning the
$D(2750)$ or $D(2760)$, which makes the $D(2600)$'s partner
assignment for the $D(2750)$ or $D(2760)$ unlikely. This conclusion
has been reached by Zhong in a chiral quark model\cite{zhong10}. We
shall discuss other possible assignments for the $D(2750)$ and
$D(2760)$ in the next subsection.

\subsection*{D. $D(2750)$ and $D(2760)$}

\indent\vspace*{-1cm}

Since the $1^-$ assignment for the $D(2750)$ or $D(2760)$ seems
unlikely as mentioned above, as shown in Table \ref{tab:assignment},
the remaining possible assignments of $D(2760)$ include the
$D(2\,^3P_0)$ and $D(1\,^3D_3)$, and the possible assignments of
$D(2750)$ are the $D(1\,^3D_3)$, $D_2(1D)$, and $D^\prime_2(1D)$.

The decay widths of $D(2760)$ as $D(2\,^3P_0)$ or $D(1\,^3D_3)$ are
listed in Table \ref{tab:pw2760}. The width of $D(1\,^3D_3)$ is
predicted to be about 73 MeV, compatible with the measured
$\Gamma(D(2760))=60.9\pm 5.6\pm 3.1$ MeV. The study in a chiral
quark model also leads to a similar result ($\sim 68$
MeV)\cite{zhong10}. The $D(1\,^3D_3)$ interpretation for the
$D(2760)$ therefore appears suitable. The width of $D(2\,^3P_0)$ is
predicted to be about 135 MeV, about 70 MeV higher than $60.9\pm
5.6\pm 3.1$ MeV. However, the lower limit of the $D(2\,^3P_0)$'s
total width is expected to be bout 72 MeV, compatible with the
measurement, which makes the $D(2\,^3P_0)$ assignments for the
$D(2750)$ also plausible.

   {\small
\begin{table}[hbt]
\begin{center}
\vspace*{-0.5cm} \caption{\label{tab:pw2760}\small Decay widths of
 $D(2760)$ as $D(2\,^3P_0)$ or $D(1\,^3D_3)$ in MeV.  A symbol``$\times$''
indicates that a decay mode is forbidden. }
 \vspace*{0.2cm}
\begin{tabular}{lcc}\hline\hline
&\multicolumn{1}{c}{$D(2\,^3P_0)$}
&\multicolumn{1}{c}{$D(1\,^3D_3)$}\\
 Mode         & $\Gamma_i$      & $\Gamma_i$\\\hline
$D^0\pi^0$        & $20.10$          & $10.74$ \\
$D^+\pi^-$        & $40.71$          & $20.92$ \\

$D_sK$            & $17.84$          & $0.82$  \\

$D\eta$           & $13.95$          & $1.77$  \\

${D^\ast}^0\pi^0$ & $\times$         & $10.42$  \\

${D^\ast}^+\pi^-$ & $\times$         & $20.29$  \\

$D^\ast\eta$      & $\times$         & $0.76$   \\

$D^\ast_s K$      & $\times$         & $0.21$   \\

$D_1(2430)^0\pi^0$& $12.25$          & $0.72$   \\

$D_1(2430)^+\pi^-$& $24.29$          & $1.41$    \\

$D_1(2420)^0\pi^0$& $2.14$          & $0.02$   \\

$D_1(2420)^+\pi^-$& $4.27$          & $0.03$   \\

$D_2(2460)^0\pi^0$& $\times$        & $0.77$   \\

$D_2(2460)^+\pi^-$& $\times$        & $1.51$    \\

$D\omega$         &  $\times$       & 0.65     \\

$D^0\rho^0$       &  $\times$       & 0.78     \\

$D^+\rho^-$       & $\times$        & 1.37  \\

                &$\Gamma_{\mbox{t}}=135.54$ &$\Gamma_{\mbox{t}}=73.17$  \\

\hline\hline
\end{tabular}
\end{center}
\end{table}
}

\begin{figure}[htb]
\begin{center}
\epsfig{file=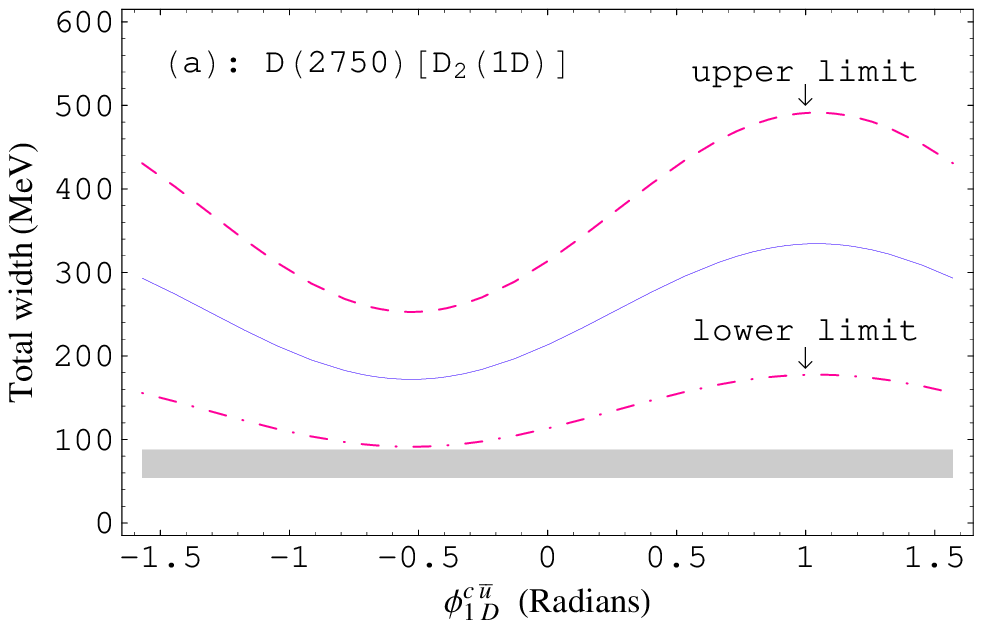,width=6.0cm, clip=}
\epsfig{file=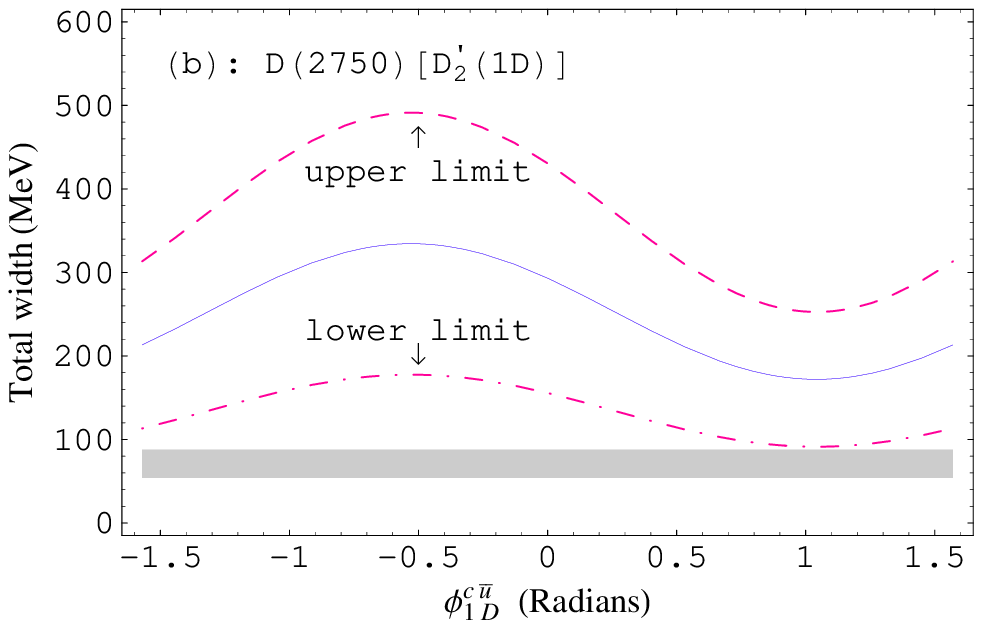,width=6.0cm, clip=}
 \vspace*{-0.3cm}
\caption{\label{fig:pw2750}\small The total width of $D(2750)$ as
the $2^-$ state versus the mixing angle. The shadow indicates the
experimentally allowed range.}
\end{center}
\end{figure}

The decay widths of $D(2750)$ as $D(1\,^3D_3)$, $D^\prime_2(1D)$,
and $D^\prime_2(1D)$ are listed in Table \ref{tab:pw2750}. The
expressions of decay widths of $D_2(1D)$ are not listed but the same
as those of $D^\prime_2(1D)$ except that the $\phi^{c\bar{u}}_{1D}$
is replaced by $\phi^{c\bar{u}}_{1D}+\pi/2$. The dependence of the
total widths of $D^\prime_2(1D)$ and $D^\prime_2(1D)$ on the mixing
angle $\phi^{c\bar{u}}_{1D}$ is illustrated in Fig.
\ref{fig:pw2750}. The total width of $D(1\,^3D_3)$ is predicted to
be about 67 MeV, consistent with the measured $\Gamma(D(2750))=71\pm
6\pm 11$ MeV, which makes the $D(1\,^3D_3)$ assignment for the
$D(2750)$ reasonable. The $D_2(1D)$ is expected to be broader than
the $D^\prime_1(2D)$. The similar behavior exists in the [$D_1(1P)$,
$D^\prime_1(1P)$]=[$D_1(2430)$, $D_1(2420)$] system where the
$D_1(1P)$ is broader than the $D^\prime_1(1P)$. From Fig.
\ref{fig:pw2750}, one can see that, at $0.697$ radians, the lower
limit of the $D_2(1D)$'s total width is substantially larger than
the upper limit of the measurement, while the lower limit of the
$D^\prime_2(1D)$'s total width is close to the upper limit of the
experiment. Therefore, if the $D(2750)$ is indeed a $2^-$ state, the
favorable quarkonium assignment would be the $D^\prime_2(1D)$ rather
than $D_2(1D)$.

   {\small
\begin{table}[hbt]
\begin{center}
\vspace*{-0.5cm} \caption{ \label{tab:pw2750}\small Decay widths of
$D(2750)$ as $D(1\,^3D_3)$, $D^\prime_2(1D)$, or $D_2(1D)$ in MeV.
$c_1\equiv\cos\phi^{c\bar{u}}_{1D}$ and
$s_1\equiv\sin\phi^{c\bar{u}}_{1D}$. Estimates of decay widths
containing $\phi^{c\bar{u}}_{1D}$ are given in terms of
$\phi^{c\bar{u}}_{1D}=0.697$ radians.
 A symbol``$\times$'' indicates
that a decay mode is forbidden.  }
 \vspace*{0.2cm}
\begin{tabular}{lcccc}\hline\hline
&\multicolumn{1}{c}{$D(1\,^3D_3)$} & \multicolumn{1}{c}{$D^\prime_2(1D)$}&\multicolumn{1}{c}{$D_2(1D)$}\\
 Mode         & $\Gamma_i$      & $\Gamma_i$   \\\hline
$D^0\pi^0$        & $10.13$         &   $\times$ & $\times$\\
$D^+\pi^-$        & $19.72$         &   $\times$ & $\times$\\

$D_sK$            & $0.71$          &   $\times$ & $\times$ \\

$D\eta$           & $1.59$          &   $\times$ & $\times$  \\

${D^\ast}^0\pi^0$   & 9.66          & $29.36c^2_1-20.33c_1s_1+25.21s^2_1=17.65$&36.93\\
${D^\ast}^+\pi^-$   & 18.81         & $58.66c^2_1-42.04c_1s_1+50.07s^2_1=34.43$&74.30\\

$D^\ast\eta$        & 0.63          & $12.67c^2_1-18.89c_1s_1+8.81s^2_1=0.27$&19.70\\

$D^\ast_s K$        & 0.16          & $0.11c^2_1+0.27c_1s_1+0.17s^2_1=1.78$&0.01\\

$D_1(2430)^0\pi^0$& $0.60$          & $0.23c^2_1-0.12c_1s_1+0.02s^2_1=0.08$&0.17\\

$D_1(2430)^+\pi^-$& $1.16$          & $0.45c^2_1-0.24c_1s_1+0.03s^2_1=0.16$&0.32\\

$D_1(2420)^0\pi^0$& $0.01$          & $1.41c^2_1+0.84c_1s_1+0.12s^2_1=1.29$&0.24\\

$D_1(2420)^+\pi^-$& $0.02$          & $2.69c^2_1+1.59c_1s_1+0.23s^2_1=2.46$&0.46\\

$D_2(2460)^0\pi^0$& 0.60            & $36.83c^2_2-60.60c_1s_1+25.26s^2_1=2.23$&59.86\\

$D_2(2460)^+\pi^-$& 1.18            & $73.99c^2_2-121.74c_1s_1+50.71s^2_1=4.48$&120.23\\

$D\omega$         & 0.47            & $18.47c^2_1+28.83c_1s_1+12.59s^2_1=30.24$&0.82\\

$D^0\rho^0$       & 0.57            & $13.35s^2_1+30.25c_1s_1+19.52c^2_1=31.87$&1.00\\

$D^+\rho^-$       & 0.99            & $37.43c^2_1+58.30c_1s_1+25.53s^2_1=61.22$&1.74\\

$D_0(2400)^0\pi^0$  &$\times$       & $0.47c^2_1+0.86c_1s_1+0.39s^2_1=0.86$&0.002\\

$D_0(2400)^+\pi^-$  &$\times$       & $0.33c^2_1+0.75c_1s_1+0.43s^2_1=0.74$&0.02\\

&$\Gamma_{\mbox{t}}=67.01$               &$\Gamma_{\mbox{t}}=293.22c^2_1-141.35c_1s_1+213.28s^2_1=189.70$&$\Gamma_{\mbox{t}}=315.81$\\

\hline\hline
\end{tabular}
\end{center}
\end{table}
}

The ratio $\frac{\Gamma(D(2760)\rightarrow
D^+\pi^-)}{\Gamma(D(2750)\rightarrow{D^\ast}^+\pi^-)}$ is
independent of the $\gamma$ and therefore is crucial to further
clarify the possible interpretations for the $D(2760)$ and
$D(2750)$. The predicted $\frac{\Gamma(D(2760)\rightarrow
D^+\pi^-)}{\Gamma(D(2750)\rightarrow{D^\ast}^+\pi^-)}$ for possible
combinations are shown in Table \ref{tab:combination}. It is obvious
that only under the identification of the [$D(2760)$, $D(2750)$] as
the [$D(1\,^3D_3)$, $D^\prime_2(1D)$] which leads to
$\frac{\Gamma(D(2760)\rightarrow
D^+\pi^-)}{\Gamma(D(2750)\rightarrow{D^\ast}^+\pi^-)}=0.6$, the
measured branching ratio of $0.42\pm 0.05\pm 0.11$ can be reasonably
accounted for. Therefore the [$D(1\,^3D_3)$, $D^\prime_2(1D)$]
assignment for the [$D(2760)$, $D(2750)$] is agreeable. The
calculations preformed by Wang in the heavy quark effective
theory\cite{wang10} also support this picture. The Babar
Collaboration suggested that $D(2760)$ and $D(2750)$ may be the
$D$-wave states, favoring our present assignment. For the
$D(1\,^3D_3)$, the main decay modes are $D\pi$, $D^\ast\pi$,
$D\rho$, $D_2(2460)K$, and $D_1(2430)K$, while for the
$D^\prime_2(1D)$ they are $D^\ast\pi$, $D\omega$, $D\rho$, and
$D_2(2460)K$. It is expected that $\Gamma(D\rho)/\Gamma(D^\ast\pi)$
for the $D(1\,^3D_3)$ and $D^\prime_2(1D)$ are about 1.8 and 0.07,
respectively. Further experimental information on these two states
in $D^\ast\pi$, $D\omega$, and $D\rho$ channels can further test the
present assignments for $D(2760)$ and $D(2750)$.

{\small
\begin{table}[hbt]
\begin{center}
\vspace*{-0.5cm} \caption{\label{tab:combination}\small
$\frac{\Gamma(D(2760)\rightarrow
D^+\pi^-)}{\Gamma(D(2750)\rightarrow{D^\ast}^+\pi^-)}$ for possible
combinations.}
 \vspace*{0.2cm}
\begin{tabular}{ccc}\hline\hline
$[D(2760),~D(2750)]$ & $\frac{\Gamma(D(2760)\rightarrow
D^+\pi^-)}{\Gamma(D(2750)\rightarrow{D^\ast}^+\pi^-)}$\\\hline
$[D(2\,^3P_0),~D(1\,^3D_3)]$                &  2.16\\

$[D(2\,^3P_0),~D^\prime_2(1D)]$                   & 1.18\\
$[D(1\,^3D_3),~D(1\,^3D_3)]$               & 1.11\\

$[D(1\,^3D_3),~D^\prime_2(1D)]$                   & 0.60\\
\hline\hline
\end{tabular}
\end{center}
\end{table}
}

The $D(2750)$ signal observed in $D^\ast\pi$ is very similar to the
$D(2760)$ signal observed in $D\pi$, their mass and width values
differ by $2.6\sigma$ and $1.5\sigma$\cite{babarcn}. Therefore, it
is likely that the $D(2760)$ and $D(2750)$ refer to the same
resonance. If so, since the $1^-$ interpretation for the $D(2760)$
or $D(2750)$ can be excluded as discussed previously, the only one
possible quarkonium assignment would be that they are the same
$D(1\,^3D_3)$, although the ratio of
$\Gamma(D^+\pi^-)/\Gamma({D^\ast}^+\pi^-)$ is somewhat larger than
the measured $0.42\pm 0.05\pm 0.11$.\footnote{Under the
$D(1\,^3D_3)$ assignment, this ratio is expected to be about 1.0 at
the initial state mass of $(2763.3+2752.4)/2$ MeV. The similar
result of about 0.9 is also obtained in a $^3P_0$ model\cite{sun10}.
The predictions from the chiral quark model\cite{zhong10} and the
heavy quark effective theory\cite{wang10} are 1.58 and 1.94,
respectively.} Clearly, the further search of the $D(2750)$ in the
$D\pi$, $D_sK$, $D\eta$, and $D_0(2400)\pi$ channels is crucial to
discriminate the $2^-$ and $3^-$ assignments for the $D(2750)$
because a $3^-$ $c\bar{n}$ state is forbidden to decay into
$D_0(2400)\pi$ while a $2^-$ $c\bar{n}$ state is forbidden to decay
into $D\pi$, $D_sK$, and $D\eta$. Also, the
$\Gamma(D\rho)/\Gamma(D^\ast\pi)$ for the $D^\prime_2(1D)$ is
remarkably different from that for the $D(1\,^3D_3)$, therefore the
experimental information on the $D(2750)$ in the $D\rho$ and
$D\omega$ is also important to differentiate these two possible
interpretations for the $D(2750)$.

In summary, if the $D(2750)$ and $D(2760)$ are confirmed to be the
same resonance, they would be the $D(1\,^3D_3)$; otherwise, the
favorable interpretation would be that the $D(2750)$ and $D(2760)$
are the $D^\prime_2(1D)$ and $D(1\,^3D_3)$, respectively. To
distinguish these two possibilities, further experimental studies on
these two states are needed. Below we turn to the charmed-strange
states.

\subsection*{E. $D_{s1}(2710)$ and $D_{sJ}(2860)$}
\indent\vspace*{-1cm}

In the 2S-1D mixing scenario, the eigenvectors of $D_{s1}(2710)$ and
its partner $D_{s1}(M_Y)$ can be written as
\begin{eqnarray}
&&|D_{s1}(2710)\rangle=\cos\theta_1 |2\,^3S_1\rangle-\sin\theta_1
|1\,^3D_1\rangle,\label{mix1}\\
&&|D_{s1}(M_Y)\rangle=\sin\theta_1 |2\,^3S_1\rangle+\cos\theta_1
|1\,^3D_1\rangle,\label{mix2}
\end{eqnarray}
where the $\theta_1$ is the $D_s(2\,^3S_1)$-$D_s(1\,^3D_1)$ mixing
angle and $M_Y$ denotes the mass of the physical state
$D_{s1}(M_Y)$.

\begin{figure}[hbt]
\begin{center}
\epsfig{file=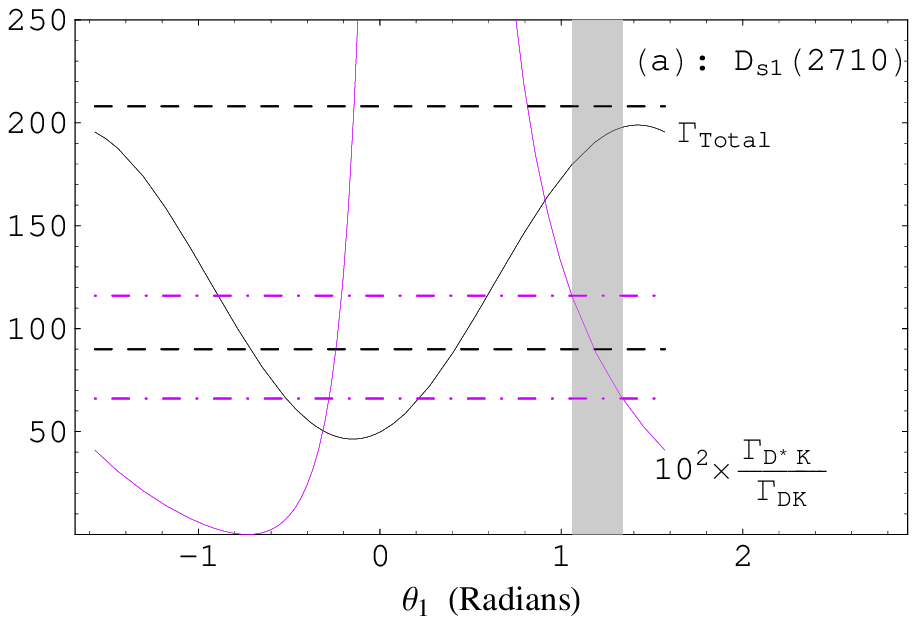,width=6.0cm, clip=}
\epsfig{file=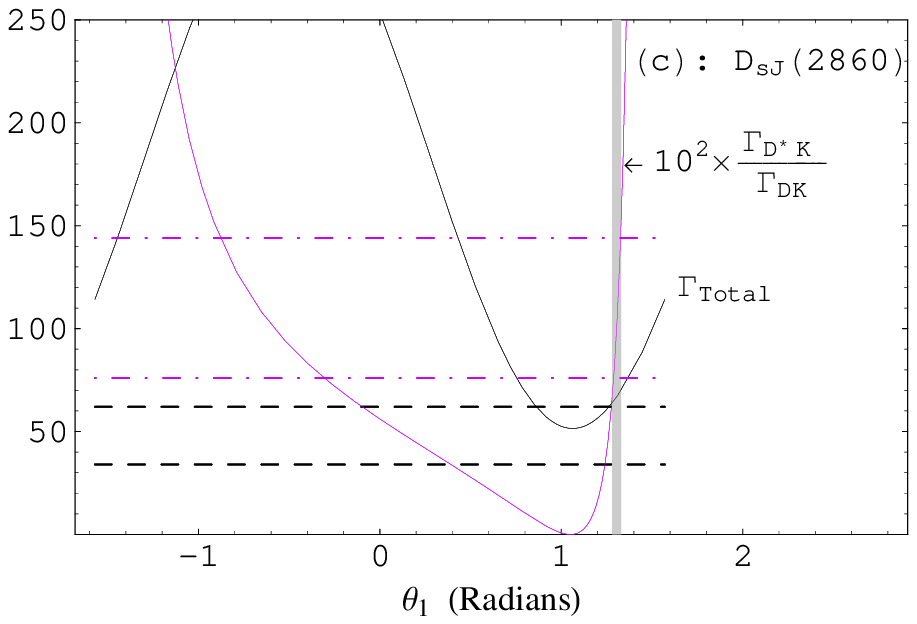,width=6.0cm, clip=}
\epsfig{file=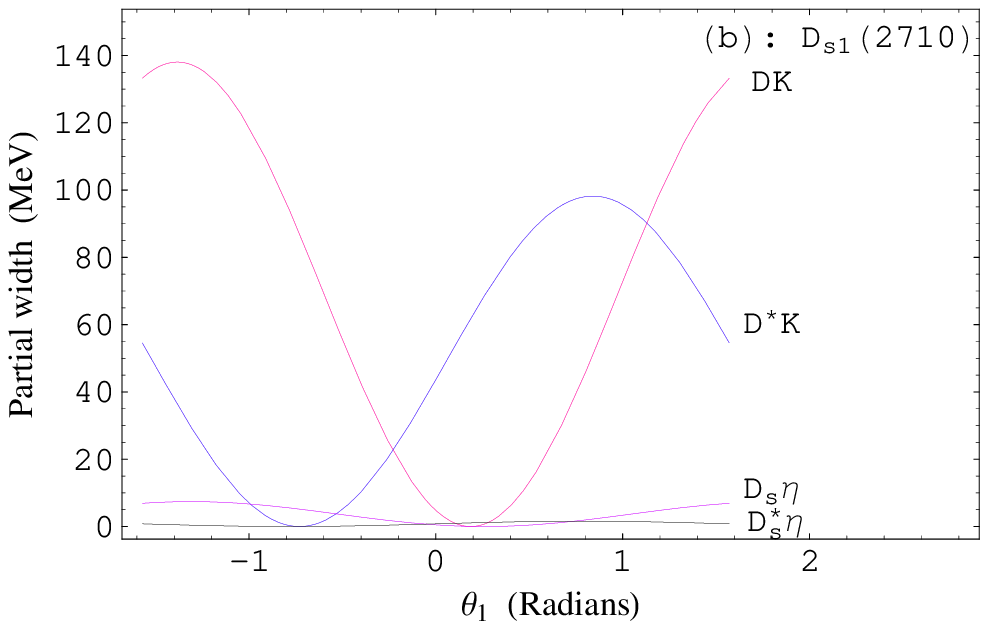,width=6.0cm,clip=}
\epsfig{file=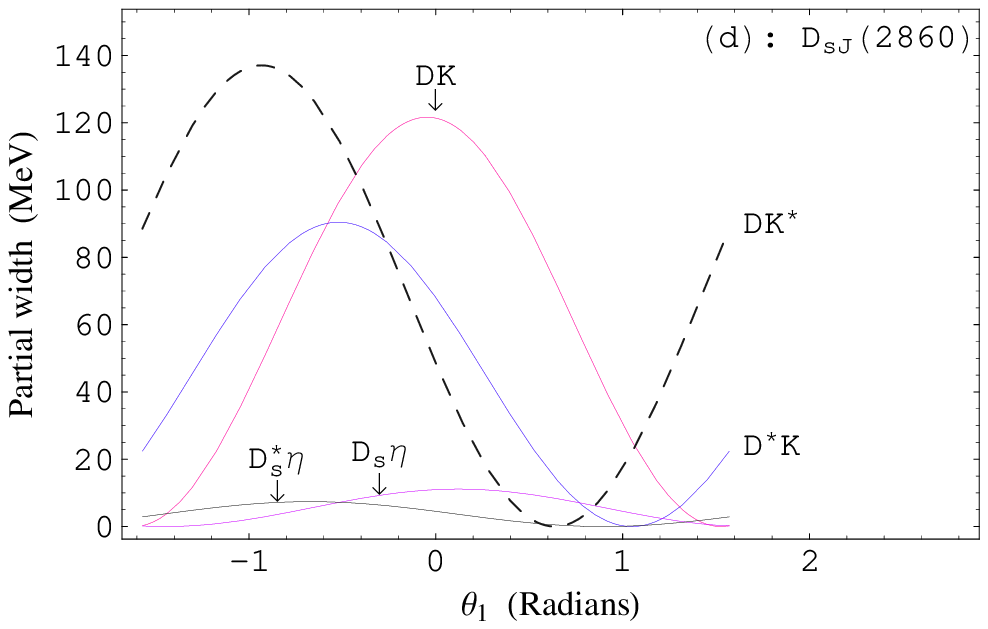,width=6.0cm,clip=}
 \vspace*{-0.3cm}
\caption{\label{fig:pw2710}\small Total widths and $\Gamma(D^\ast
K)/\Gamma(DK)$ of $D_{s1}(2710)$ and $D_{sJ}(2860)$ versus
$\theta_1$. The horizontal dashed lines indicate the upper and lower
limits of experimental data.  }
\end{center}
\end{figure}

The decay widths of $D_{s1}(2710)$ are listed in Table
\ref{tab:pw2710}.  The variations of decay widths and $\Gamma(D^\ast
K)/\Gamma(DK)$ with the mixing angle $\theta_1$ are illustrated in
Fig. \ref{fig:pw2710}. Clearly, with $1.06 \leq \theta_1 \leq 1.34$
radians, both the total width and $\Gamma(D^\ast K)/\Gamma(DK)$ of
 $D_{s1}(2710)$ can be well reproduced (see Fig. \ref{fig:pw2710}
(a)). Also, in this mixing angle range, the main decay modes are
$DK$ and $D^\ast K$ (see Fig. \ref{fig:pw2710} (b)), in accord with
the observation of the $D_{s1}(2710)$ in the $DK$ and $D^\ast K$.
Therefore, the picture of $D_{s1}(2710)$ being in fact a mixture of
the $D_s(2\,^3S_1)$ and $D_s(1\,^3D_1)$ seems convincing. The
studies in a chiral quark model\cite{zhong09} and a $^3P_0$
model\cite{li09} also favor this interpretation. Future experimental
information on the $D_{s1}(2710)$ in the remaining channels
$D_s\eta$ and $D^\ast_s\eta$ can provide a consistent check for this
assignment.

{\small
\begin{table}[hbt]
\begin{center}
\vspace*{-0.5cm} \caption{\label{tab:pw2710}\small Decay widths of
 $D_{s1}(2710)$ and $D_{sJ}(2860)$ as the $1^-$ states.
 $c_2\equiv\cos\theta_1$, and
$s_2\equiv\sin\theta_1$. A dash indicates that a decay mode is below
threshold.}
 \vspace*{0.2cm}
\begin{tabular}{lcc}\hline\hline
                  & \multicolumn{1}{c}{$D_{s1}(2710)$}&\multicolumn{1}{c}{$D_{sJ}(2860)$}\\
 Mode             & $\Gamma_i$                      &$\Gamma_i$\\\hline
$D^0K^+$          & $2.29c^2_2-24.74c_2s_2+66.71s^2_2$      &$60.22c^2_2-6.47c_2s_2+0.17s^2_2$   \\
$D^+K^0$          & $2.49c^2_2-25.79c_2s_2+66.54s^2_2$      &$61.17c^2_2-4.89c_2s_2+0.10s^2_2$   \\
$D_s\eta$         & $0.49c^2_2-3.69c_2s_2+6.90s^2_2$        &$10.96c^2_2+2.64c_2s_2+0.16s^2_2$    \\
${D^\ast}^0 K^+$  & $21.81c^2_2+49.05c_2s_2+27.57s^2_2$     &$33.93c^2_2-38.36c_2s_2+10.84s^2_2$ \\
${D^\ast}^+ K^0$  & $21.83c^2_2+48.55c_2s_2+26.99s^2_2$     &$34.15c^2_2-36.69c_2s_2+11.53s^2_2$ \\
$D^\ast_s\eta$    & $0.79c^2_2+1.59c_2s_2+0.80s^2_2$        &$4.51c^2_2-7.20c_2s_2+2.87s^2_2$\\
$D^0{K^\ast}^+$   & $-$                                & $25.22c^2_2-67.82c_2s_2+45.58s^2_2$\\
$D^+{K^\ast}^0$   & $-$                                & $23.32c^2_2-63.30c_2s_2+42.94s^2_2$\\
                  &$\Gamma_{\mbox{t}}=49.72c^2_2+44.95c_2s_2+195.52s^2_2$
                  &$\Gamma_{\mbox{t}}=253.49c^2_2-225.07c_2s_2+114.20s^2_2$\\\hline\hline
\end{tabular}
\end{center}
\end{table}
}

\begin{figure}[hbt]
\begin{center}
\epsfig{file=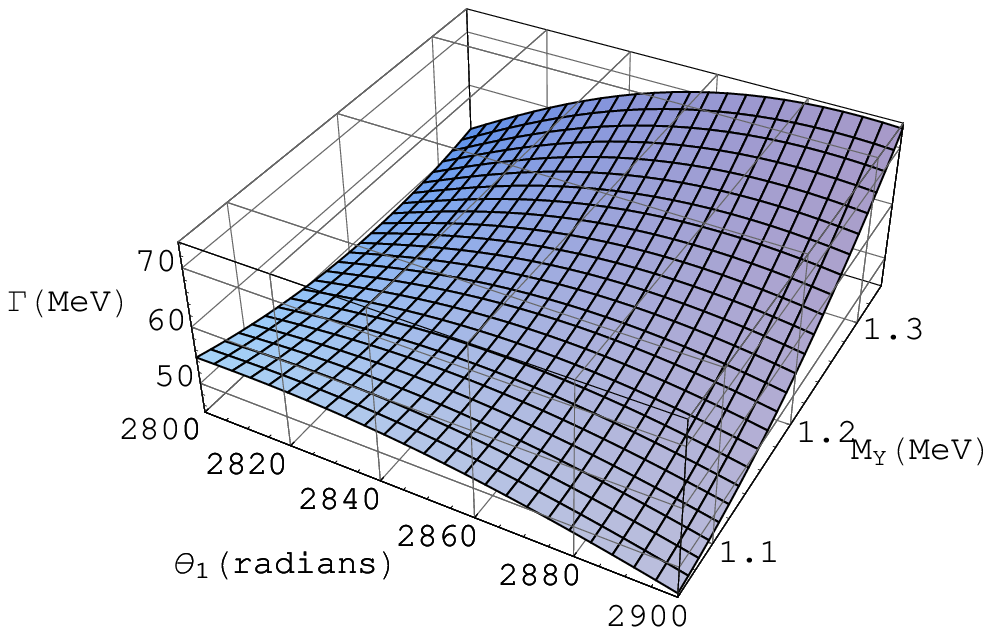,width=6.0cm, clip=}
\epsfig{file=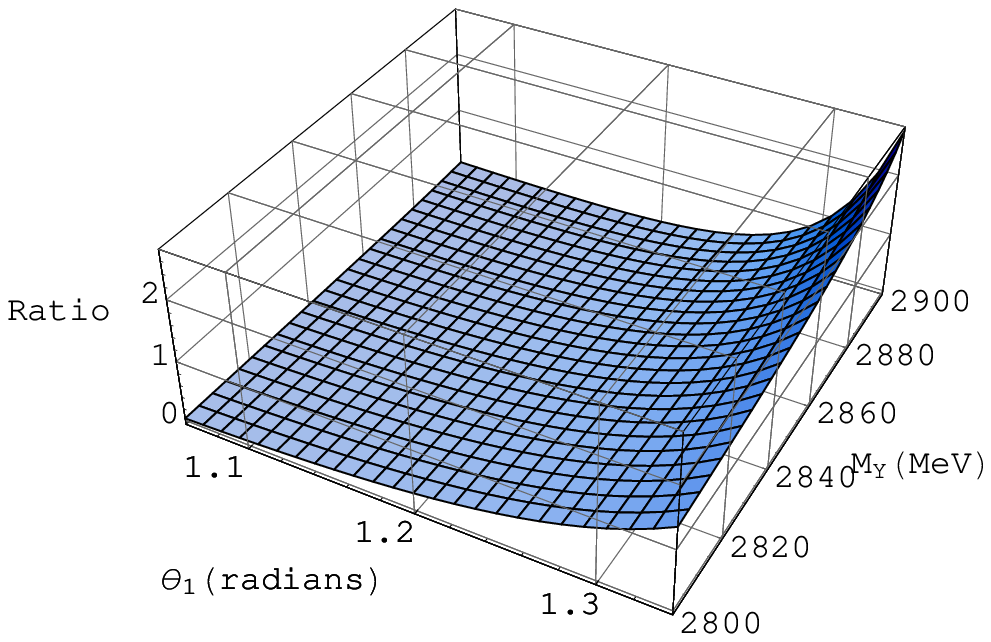,width=6.0cm, clip=}

 \vspace*{-0.3cm}
\caption{\label{fig:predcs}\small Total width and branching ratio
$\Gamma(D^\ast K)/\Gamma(DK)$ of $D_{s1}(M_X)$ versus $M_Y$ and
$\theta_1$.}
\end{center}
\end{figure}

According to $M^2_Y=
M^2_{D_s(2\,^3S_1)}+M^2_{D_s(1\,^3D_1)}-M^2_{D_{s1}(2710)}$, the
$D(M_Y)$ is expected to have a mass of about 2.81 GeV. Other
approaches predicted $M_Y\simeq 2.8\sim 2.9$ GeV (see Table
\ref{tab:mesonmass1}). The total width and the branching ratio
$\Gamma(D^\ast K)/\Gamma(DK)$ for the $D_{s1}(M_Y)$ as functions of
the initial state mass $M_Y$ and the mixing angle $\theta_1$ are
illustrated in Fig. \ref{fig:predcs}. The $M_Y$ is restricted to be
$2800\sim 2900$ MeV and the $\theta_1$ is restricted to be $1.06\sim
1.34$ radians. With the variations of the initial state mass and the
mixing angle, the total width of $D_{s1}(M_Y)$ varies from about 40
to 70 MeV and the $\Gamma(D^\ast K)/\Gamma(DK)$ varies from about
0.04 to 2.71.

Both the predicted mass and width of $D_{s1}(M_Y)$ are consistent
with those of $D_{sJ}(2860)$. Therefore, if $D_{sJ}(2860)$ is a
$1^-$ state,
  it would be a natural candidate for the $D_{s1}(M_Y)$. Under this
  picture, the numerical results for the decays widths of $D_{sJ}(2860)$ are
listed in Table \ref{tab:pw2710}, and the dependence of the total
width as well as $\Gamma(D^\ast K)/\Gamma(DK)$ on the mixing angle
$\theta_1$ is illustrated in Figs.\ref{fig:pw2710}. It is clear that
both the total width and $\Gamma(D^\ast K)/\Gamma(DK)$ of
$D_{sJ}(2860)$ can be well reproduced with $1.29\leq \theta_1\leq
1.33$ radians, just lying on the range of $1.06\leq \theta_1\leq
1.34$ radians (see \ref{fig:pw2710}(c)). Also, in this mixing angle
range, the main decay modes are $DK^\ast$, $D^\ast K$ and $DK$,
consistent with the observation of the $D_{sJ}(2860)$ in $DK$ and
$D^\ast K$. Therefore, the identification of $D_{sJ}(2860)$ as the
partner of $D_{s1}(2710)$ appears convincing. The study in a $^3P_0$
model\cite{li09} also favors this assignment.

The $D_{sJ}(2860)$ could also be the $D_s(1\,^3D_3)$ as shown in
Table \ref{tab:assignment}. In this case, the decay widths are
listed in Table \ref{tab:pw2860}. It is expected that $\Gamma\simeq
77$ MeV and branching ratio $\Gamma(D^\ast K)/\Gamma(DK)\simeq 0.8$.
The predicted $\Gamma(D^\ast K)/\Gamma(DK)\simeq 0.8$ is consistent
with the measurement. The predicted total with is about 20 MeV
larger than the experiment. The lower limit of the predicted width
is about 41 MeV, consistent with the experiment. The main decay
modes are $DK$ and $D^\ast K$, in accord with the observation. The
$D_s(1\,^3D_3)$ interpretation for the $D_{sJ}(2860)$ thus seems
likely. This assignment is also favored by the studies in the
$^3P_0$ model\cite{li09} and lattice QCD\cite{lattice}.

 {\small
\begin{table}[hbt]
\begin{center}
\vspace*{-0.5cm} \caption{\label{tab:pw2860}\small Decay widths of
$D_{sJ}(2860)$ as $D_s(1\,^3D_3)$ in MeV. }
 \vspace*{0.2cm}
\begin{tabular}{cccccccccc}\hline\hline
$D^0K^+$ & $D^+K^0$ &$D_s\eta$ & ${D^\ast}^0 K^+$
             &${D^\ast}^+ K^0$ & $D^\ast_s\eta$&
             $D^0{K^\ast}^+$&$D^+{K^\ast}^0$&\mbox{Total}
               \\\hline
  21.38&
20.53&0.93&16.18&15.40&0.35&1.31&0.99&$77.05$\\\hline\hline
\end{tabular}
\end{center}
\end{table}
}

Both the $1^-$ and $3^-$ interpretations for the $D_{sJ}(2860)$
appear reasonable. The available experimental information on the
$D_{sJ}(2860)$ is not enough to discriminate these two
possibilities. However, the differences between the $1^-$ and $3^-$
interpretations are evident. For example, for the $3^-$ assignment,
$\Gamma(DK)/\Gamma(D^\ast_s\eta)\simeq 119.7$,
$\Gamma(DK)/\Gamma(D_s\eta)\simeq 45.1$, and
$\Gamma(DK)/\Gamma(DK^\ast)\simeq 18.2$,  while for the $1^-$
assignment, $\Gamma(DK)/\Gamma(D^\ast_s\eta)\simeq 3.5\sim 6.1$,
$\Gamma(DK)/\Gamma(D_s\eta)\simeq 3.3\sim 3.9$, and
$\Gamma(DK)/\Gamma(DK^\ast)\simeq 0.08\sim 0.13$. Further
experimental information on the $D_{sJ}(2860)$ in $D_s\eta$,
$D^\ast_s\eta$, and $DK^\ast$ channels is crucial to distinguish
these two possible assignments.

\subsection*{F. $D_{sJ}(3040)$}
\indent\vspace*{-1cm}

\begin{figure}[htb]
\begin{center}
\epsfig{file=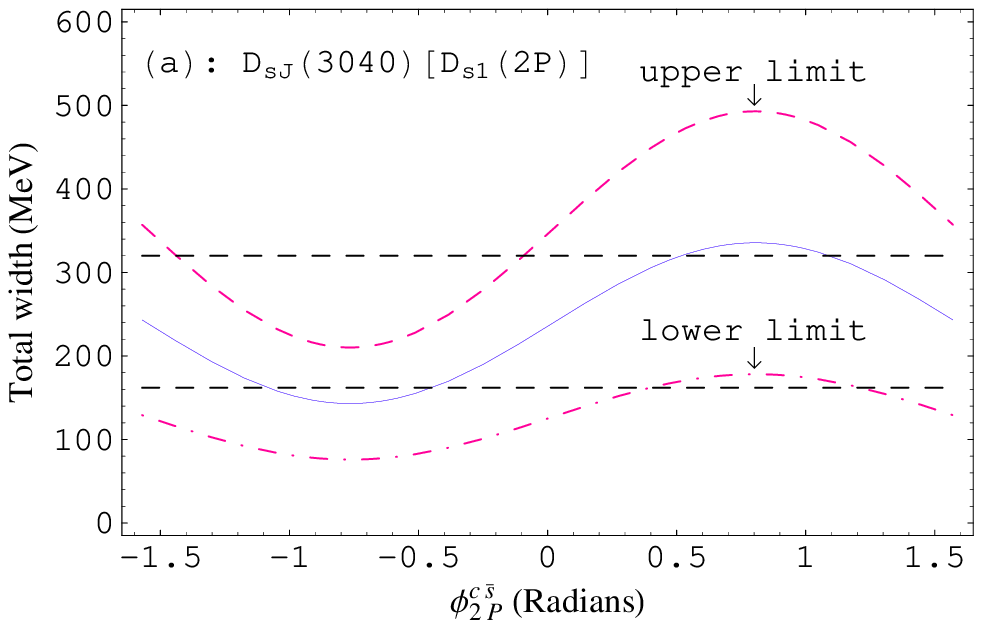,width=6.0cm, clip=}
\epsfig{file=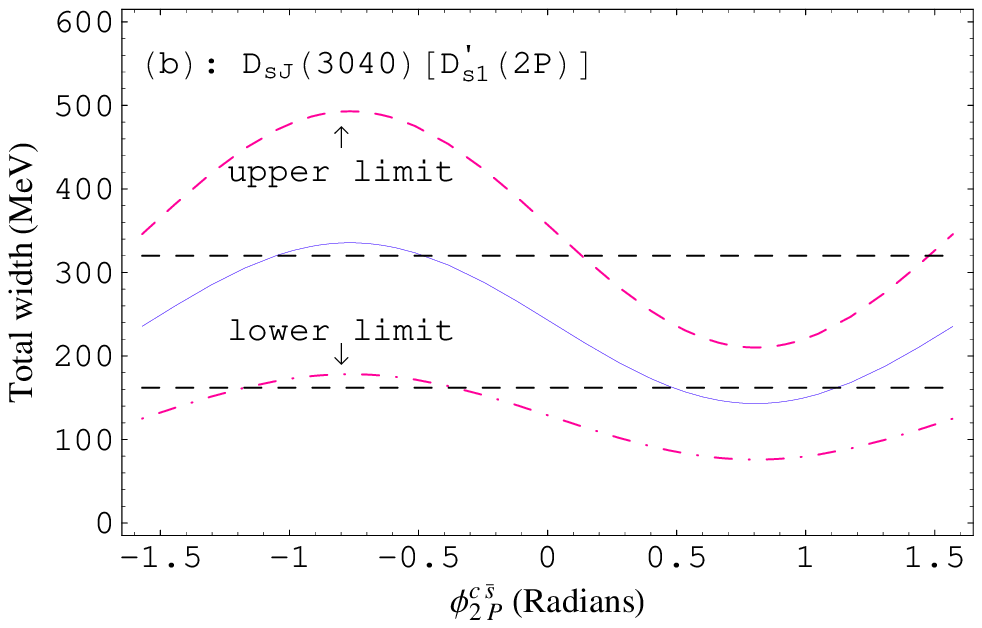,width=6.0cm, clip=}
 \vspace*{-0.3cm}
\caption{\label{fig:pw3040}\small The total width of $D_{sJ}(3040)$
as the $1^+$ state versus the mixing angle. The horizontal dashed
lines indicate the upper and lower limits of experimental data.}
\end{center}
\end{figure}

\begin{figure}[htb]
\begin{center}
\epsfig{file=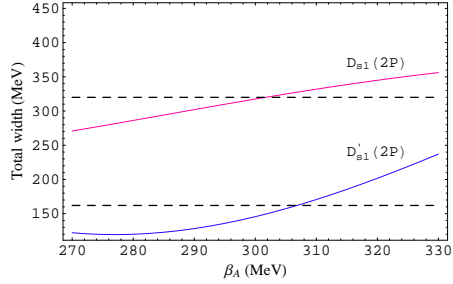,width=6.0cm, clip=}

 \vspace*{-0.3cm}
\caption{\label{fig:pw30401}\small The total width of $D_{sJ}(3040)$
as the $1^+$ state versus the $\beta_A$. Here, $\beta_{2\,^3P_1}$ is
set to $\beta_{2\,^1P_1}$. The horizontal dashed lines indicate the
upper and lower limits of experimental data.}
\end{center}
\end{figure}

The decay widths of $D_{sJ}(3040)$ as $D_{s1}(2P)$ or
$D^\prime_{s1}(2P)$ are listed in Table \ref{tab:pw3040}. The
expressions of the decay widths of
 $D^\prime_{s1}(2P)$ are not listed but the same as those of the
$D_{s1}(2P)$ except that the $\phi^{c\bar{s}}_{2P}$ is replaced by
$\phi^{c\bar{s}}_{2P}+\pi/2$. The dependence, of the total width of
$D_{sJ}(3040)$ as $1^+$ state, on the mixing angle
$\phi^{c\bar{s}}_{2P}$ and $\beta_A$ are illustrated in Fig.
\ref{fig:pw3040} and Fig. \ref{fig:pw30401}.

 As can be seen in Fig.
\ref{fig:pw3040}, at $\phi^{c\bar{s}}_{2P}=0.564$ radians, the
predicted $\Gamma(D_{s1}(2P))$ and $\Gamma(D^\prime_{s1}(2P))$ are
close to the upper and lower limits of the experimental data of
$239\pm 35 ^{+46}_{-42}$ MeV, respectively. The similar behavior
also exists at about $\beta_A=306$ MeV, as shown in Fig.
\ref{fig:pw30401}. Within the theoretical and experimental errors,
the predicted total widths for both $D_{s1}(2P)$ and
$D^\prime_{s1}(2P)$ are comparable with the experiment. Therefore,
both the $D_{s1}(2P)$ and $D^\prime_{s1}(2P)$ assignments for the
$D_{sJ}(3040)$ seem likely based on its measured total width. It
should be noted that since the experimental errors of
$\Gamma(D_{sJ}(3040)$ is large, the improved measurement of
$\Gamma(D_{sJ}(3040)$ is needed to confirm our present assignment.

Also, Fig. \ref{fig:pw30401} indicates that in the vicinity of
initial state $\beta=300$ ($270\leq \beta_A \leq 330$ MeV), the
$D_{s1}(2P)$ is expected to be about $100\sim 150$ MeV broader than
the $D^\prime_{s1}(2P)$ in width, which is consistent with the
prediction from the heavy quark effective theory. In the framework
of the heavy quark effective theory, the $D_{s1}(2P)$ is the $1^+$
state existing in $S=(0^+,1^+)$ doublet while $D^\prime_{s1}(2P)$
corresponds to the $1^+$ state of $T=(1^+,2^+)$ doublet, and the
$1^+$ state of $S$ doublet is predicted to be broader than the one
of $T$ doublet\cite{sun09,liuxiang}. The similar conclusion has been
reached in calculations from the $^3P_0$ model\cite{sun09} and the
chiral quark model\cite{zhong09}.

For the $D_{s1}(2P)$, the main decay modes are $D_2(2460)K$, $D^\ast
K$, $D^\ast K^\ast$, $DK^\ast$, and $D_0(2400)K$, while for the
$D^\prime_{s1}(2P)$ they are $D^\ast K^\ast$, $DK^\ast$, $D^\ast K$,
$D_0(2400)K$, $D_1(2420)K$, and $D_2(2460)K$. The branching ratios
$\Gamma(D^\ast K^\ast)/\Gamma(D^\ast K)$, $\Gamma(D
K^\ast)/\Gamma(D^\ast K)$ and $\Gamma(D_2(2460)K)/\Gamma(D^\ast K)$
are expected to be respectively about 0.9, 0.8, and 1.1 for the
$D_{s1}(2P)$, while about 5.0, 2.0, and 0.4 for the
$D^\prime_{s1}(2P)$. The decay patterns for these two assignments
are different. The additional experimental information on the
branching ratios of $D_{sJ}(3040)$ are important to discriminate
these two possibilities.

 {\small
\begin{table}[hbt]
\begin{center}
\vspace*{-0.5cm} \caption{\label{tab:pw3040}\small Decay widths of
$D_{sJ}(3040)$ as the $D_{s1}(2P)$ or $D^\prime_{s1}(2P)$ in MeV.
$c_3\equiv\cos\phi^{c\bar{s}}_{2P}$,
$s_3\equiv\sin\phi^{c\bar{s}}_{2P}$. Estimates of decay widths
containing $\phi^{c\bar{s}}_{2P}$ are given in terms of
$\phi^{c\bar{s}}_{2P}=0.564$ radians. }
 \vspace*{0.2cm}
\begin{tabular}{lcc}\hline\hline
           & \multicolumn{1}{c}{$D_{s1}(2P)$}&\multicolumn{1}{c}{$D^\prime_{s1}(2P)$}\\
 Mode                                  &$\Gamma_i$&$\Gamma_i$\\\hline

$D^0{K^\ast}^+$                        & $25.33c^2_3+18.27c_3s_3+19.25s^2_3=31.85$&12.74\\
$D^+{K^\ast}^0$                        & $24.51c^2_3+16.28c_3s_3+19.08s^2_3=30.31$&13.27\\

$D_s\phi$                                  &$0.45c^2_3-0.05c_3s_3+0.34s^2_3=0.39$&0.40    \\

${D^\ast}^0 K^+$                              &$14.92c^2_3+35.09c_3s_3+27.42s^2_3=34.35$&7.99 \\
${D^\ast}^+ K^0$                              &$14.91c^2_3+36.01c_3s_3+27.72s^2_3=34.84$&7.79 \\
$D^\ast_s\eta$                                &$2.39c^2_3+6.84c_3s_3+4.90s^2_3=6.20$&1.10\\

${D^\ast}^0 {K^\ast}^+$                 &$31.08c^2_3+44.35s^2_3=34.59$&39.84 \\
${D^\ast}^+ {K^\ast}^0$                 &$28.83c^2_3+40.77s^2_3=32.24$&37.36 \\

$D_1(2430)^0K^+$                  &$0.01c^2_3-0.21c_3s_3+2.08s^2_3=0.50$&1.59\\
$D_1(2430)^+K^0$                  &$0.01c^2_3-0.21c_3s_3+1.99s^2_3=0.48$&1.52\\
$D_1(2420)^0K^+$                 &$0.01c^2_3-0.37c_3s_3+6.74s^2_3=1.76$&4.99\\
$D_1(2420)^+K^0$                 &$0.01c^2_3-0.38c_3s_3+6.77s^2_3=1.77$&5.01\\
$D_2(2460)^0K^+$                &$21.22c^2_3+40.71c_3s_3+21.47s^2_3=39.68$&3.00\\
$D_2(2460)^+K^0$                 &$20.79c^2_3+40.05c_3s_3+21.07s^2_3=38.97$&2.89\\

$D_0(2400)^0K^+$                     &$25.92c^2_3+1.23c_3s_3+0.01s^2_3=19.07$&6.86\\
$D_0(2400)^+K^0$                     &$20.81c^2_3-1.05c_3s_3+0.01s^2_3=14.39$&6.43\\

$D_{s1}(2460)\eta$               &$(0.05c^2_3-1.54c_3s_3+12.89s^2_3)\times 10^{-2}=0.03$&0.10\\
$D_{s0}(2317)\eta$                   &$(431.34c^2_3+8.87c_3s_3+0.05s^2_3)\times 10^{-2}=3.12$&1.19\\

                  &$\Gamma_{\mbox{t}}=235.50c^2_3+192.29c_3s_3+243.11s^2_3=324.55$&$\Gamma_{\mbox{t}}=154.06$\\\hline\hline
\end{tabular}
\end{center}
\end{table}
}


\section*{IV. Radiative decays}
\indent\vspace*{-1cm}

It is well known that radiative transitions can probe the internal
charge structure of hadrons, and therefore they will likely play an
important role in determining the quantum numbers and hadronic
structures of these new open-charm mesons. In this section, we shall
evaluate the $E1$ and $M1$ transitions widths of these open-charm
states.

The partial widths for the $E1$ and $M1$ transitions between the
$v=n~^{2S+1}L_J$ and
$v^\prime=n^\prime\,^{2S^\prime+1}L^\prime_{J^\prime}$ $c\bar{q}$
states in the nonrelativistic quark model are given
by\cite{closeq,e1}
\begin{eqnarray}
&&\Gamma_{E1}(v\rightarrow v^\prime+\gamma)=\frac{4\alpha e^2_Q}{3}
C_{fi}\delta_{SS^\prime}\left |\langle v^\prime|r|v\rangle \right
|^2\frac{E^3_\gamma E_f}{M_i}
\\
&&\Gamma_{M1}(v\rightarrow v^\prime+\gamma)=\frac{\alpha e^{\prime
2}_Q
}{3}\frac{2J^\prime+1}{2L+1}\delta_{LL^\prime}\delta_{SS^\prime\pm
1}\left |\langle v^\prime|j_0\left(\frac{E_\gamma
r}{2}\right)|v\rangle \right |^2\frac{E^3_\gamma E_f}{M_i} ,
\end{eqnarray}
where $e_Q=\frac{m_{q}Q_c+m_cQ_q}{(m_q+m_c)}$,
$e^\prime_Q=\frac{m_{q}Q_c+m_cQ_q}{(m_qm_c)}$, $Q_c$ and $Q_q$
denote the quark $c$ and $q$ charges in units of $|e|$,
respectively. $\alpha=1/137$ is the fine-structure constant,
$E_\gamma$ is the final photon energy, $E_f$ is the energy of the
final state $n^\prime\,^{2S^\prime+1}L^\prime_{J^\prime}$, $M_i$ is
the initial state mass, and the angular matrix element $C_{fi}$ is
\begin{eqnarray}
C_{fi}=\mbox{Max}(L,L^\prime)(2J^\prime+1)\left\{\begin{array}{ccc}L^\prime&J^\prime&S\\
J&L&1\end{array}\right\}^2. \label{cfi}
\end{eqnarray}

The wave functions used to evaluate the matrix element $\langle
v^\prime|r|v\rangle$ and $\langle v^\prime|j_0(E_\gamma
r/2)|v\rangle$ are obtained from the nonrelativistic quark model
(\ref{cqm}). According to the PDG\cite{pdg08}, the well established
$D$ and $D_s$ states include the $D$, $D_s$, $D^\ast$, $D^\ast_s$,
$D_0(2400)$, $D_{s0}(2317)$, $D_1(2430)$, $D_1(2420)$,
$D_{s1}(2460)$, $D_{s1}(2536)$, $D_2(2460)$, and $D_{s2}(2573)$.
Therefore, we only consider the processes where the final states
contain the ground $S$ and $P$-wave open-charm mesons. The resulting
$E1$ transitions widths of these open-charm states for the favorable
assignments mentioned in above sections together with the photon
energies are given in Tables
\ref{tab:rad25502600}-\ref{tab:rad3040}. The $M1$ transitions widths
are given in Table \ref{tab:radm1}.

As can be seen in Table \ref{tab:rad2760}, the $D_1(2420)\gamma$ and
$D_1(2430)\gamma$ are clearly of great interest to discriminate the
$2^-$ and $3^-$ interpretations for the $D(2750)$, since these modes
are forbidden for a $3^-$ state while allowable for a $2^-$ state.
Especially, the $\Gamma(D^\prime_2(1D)\rightarrow D_1(2420)\gamma)$
is expected to be about 757 keV and thus becomes an experimentally
promising process.

Similarly, from Table \ref{tab:rad2710}, the experimental
information on the $D_{sJ}(2860)$ in the $D_{s0}(2317)\gamma$,
$D_{s1}(2459)\gamma$, and $D_{s1}(2535)\gamma$ would be important
to discriminate the $1^-$ and $3^-$ interpretations since these
decay modes are forbidden for the $3^-$ state while allowable for
the $1^-$ state.

As for the $M1$ transitions, experimental study on the ratio $R=
\frac{{\cal{B}}(D_{sJ}(3040)\rightarrow
D_{s1}(2460))}{{\cal{B}}(D_{sJ}(3040)\rightarrow D_{s1}(2535))}$
would be useful to discriminate the $D_{s1}(2P)$ and
$D^\prime_{s1}(2P)$ interpretations since it is expected that
$R(D_{s1}(2P))\approx 11.7$ while $R(D^\prime_{s1}(2P))\approx 1.3$.

{\small
\begin{table}[hbt]
\begin{center}
\vspace*{-0.5cm} \caption{\label{tab:rad25502600}\small $E1$
transitions widths of $D(2550)$ and $D(2600)$. $E_\gamma$ in MeV,
$\Gamma$ in keV, $c\equiv\cos\theta$, and $s\equiv\sin\theta$.
Estimates of decay widths containing $\theta$ are given in terms of
$\theta=0.4$ radians. A symbol``$\times$'' indicates that a decay
mode is forbidden. }
 \vspace*{0.2cm}
\begin{tabular}{lcccc}\hline\hline
      & \multicolumn{2}{c}{$D(2550)$[$2^1S_0$]}&\multicolumn{2}{c}{$D(2600)$[2S-1D]}\\
Final meson &$E_\gamma$& $\Gamma$& $E_\gamma$& $\Gamma$\\\hline

$D_{2}(2460)^0$ &$\times$ &$\times$ &
143&$86.4c^2+32.7cs+3.1s^2\simeq 85.5 $\\

$D_0(2400)^0$ &$\times$ &$\times$ &
283&$125.8c^2+475.6cs+449.6s^2\simeq 345.5$\\

$D_1(2430)^0$ &110 & $76.0$&175
&$11.8c^2+22.3cs+10.6s^2\simeq 19.6$\\

$D_1(2420)^0$ &114 & $12.3$& 180&$166.2c^2+87.9cs+78.6s^2\simeq
184.3$\\\hline\hline
      \end{tabular}
\end{center}
\end{table}
}

{\small
\begin{table}[hbt]
\begin{center}
\vspace*{-0.5cm} \caption{\label{tab:rad2760}\small $E1$ transitions
widths of $D(2760)$ and $D(2750)$. $E_\gamma$ in MeV, $\Gamma$ in
keV, $c_1\equiv\cos\phi^{c\bar{u}}_{2D}$, and
$s_1\equiv\sin\phi^{c\bar{u}}_{1D}$. Estimates of decay widths
containing $\phi^{c\bar{u}}_{1D}$ are given in terms of
$\phi^{c\bar{u}}_{1D}=0.697$ radians. A symbol``$\times$'' indicates
that a decay mode is forbidden.}
 \vspace*{0.2cm}
\begin{tabular}{lccccc}\hline\hline
      & \multicolumn{3}{c}{$D(2750)$}&\multicolumn{2}{c}{$D(2760)$}\\
Final meson &$E_\gamma$& $\Gamma$[$D^\prime_2(1D)$]&
$\Gamma$[$1^3D_3$]& $E_\gamma$& $\Gamma$[$1^3D_3$]\\\hline

$D_2(2460)^0$ &276 & $188.5c^2_1\simeq110.8$
&753.9&286 & 834.2\\

$D_0(2400)^0$  & $\times$&$\times$
&$\times$&$\times$ & $\times$\\

$D_{1}(2430)^0$ &306 &
$96.2c^2_1-578.5c_1s_1+869.4s^2_1\simeq130.1$& $\times$
&$\times$ & $\times$\\

$D_{1}(2420)^0$ &310 &
$693.4c^2_1-601.3c_1s_1+130.4s^2_1\simeq757.3$& $\times$&$\times$ &
$\times$\\\hline\hline
      \end{tabular}
\end{center}
\end{table}
}

{\small
\begin{table}[hbt]
\begin{center}
\vspace*{-0.5cm} \caption{ \label{tab:rad2710}\small $E1$
transitions widths of $D_{s1}(2710)$ and $D_{sJ}(2860)$. $E_\gamma$
in MeV, $\Gamma$ in keV, $c_2\equiv\cos\theta_1$, and
$s_2\equiv\sin\theta_1$. Estimates of decay widths containing
$\theta_1$ are given in terms of $\theta_1=1.31$ radians.
 A symbol``$\times$'' indicates that a
decay mode is forbidden.}
 \vspace*{0.2cm}
\begin{tabular}{lccccc}\hline\hline
      & \multicolumn{2}{c}{$D_{s1}(2710)$}&\multicolumn{3}{c}{$D_{sJ}(2860)$}\\
Final meson &$E_\gamma$& $\Gamma$[2S-1D]& $E_\gamma$&
$\Gamma$[2S-1D]&$\Gamma$[$1\,^3D_3$]\\\hline

$D_{s2}(2573)$ &134 & $0.4c^2_2+0.2c_2s_2+0.02s^2_2\simeq 0.1$&
275&$0.1c^2_2-1.4c_2s_2+3.7s^2_2\simeq 3.1$&4.6\\

$D_{s0}(2317)$ &364 & $1.6c^2_2+6.1c_2s_2+5.7s^2_2\simeq 6.9$&
492&$13.6c^2_2-14.5c_2s_2+3.9s^2_2\simeq 0.9$&$\times$\\

$D_{s1}(2460)$ &239 & $0.3c^2_2+0.5c_2s_2+0.2s^2_2\simeq 0.3$&374
&$0.8c^2_2-1.7c_2s_2+0.9s^2_2\simeq 0.5$&$\times$\\

$D_{s1}(2536)$ &169 & $0.4c^2_2+0.8c_2s_2+0.4s^2_2\simeq 0.6$&
308&$2.2c^2_2-4.7c_2s_2+2.5s^2_2\simeq 1.3$&$\times$\\\hline\hline
      \end{tabular}
\end{center}
\end{table}
}

{\small
\begin{table}[hbt]
\begin{center}
\vspace*{-0.5cm} \caption{ \label{tab:rad3040}\small $E1$
transitions widths of $D_{sJ}(3040)$. $E_\gamma$ in MeV, $\Gamma$ in
keV, $c_3\equiv\cos\phi^{c\bar{s}}_{2P}$, and
$s_3\equiv\sin\phi^{c\bar{s}}_{2P}$. Estimates of decay widths
containing $\phi^{c\bar{s}}_{2P}$ are given in terms of
$\phi^{c\bar{s}}_{2P}=0.564$ radians.}
 \vspace*{0.2cm}
\begin{tabular}{lccc}\hline\hline
      & \multicolumn{3}{c}{$D_{sJ}(3040)$}\\
Final meson &$E_\gamma$& $\Gamma$[$D_{s1}(2P)$]&
$\Gamma$[$D^\prime_{s1}(2P)$]\\\hline

$D_s(1969)$ &886 & $1.6c^2_3\simeq1.1$
& $1.6s^2_3\simeq0.5$\\

$D^\ast_s(2112)$ &789 & $0.5s^2_3\simeq0.1$
& $0.5c^2_3\simeq0.4$\\

$D_{s2}(2573)$ &435 & $0.6s^2_3\simeq0.2$
&$0.6c^2_3\simeq0.4$\\

$D_{s0}(2317)$ &640 & $1.5s^2_3\simeq0.4$
& $1.5c^2_3\simeq1.1$\\

$D_{s1}(2460)$ &528 & $2.1c^2_3+1.0c_3s_3+0.1s^2_3\simeq1.9$
& $2.1s^2_3-1.0c_3s_3+0.1c^2_3\simeq0.2$\\

$D_{s1}(2536)$ &466 & $0.3c^2_3-0.7c_3s_3+0.4s^2_3\simeq0.01$ &
$0.3s^2_3+0.7c_3s_3+0.4c^2_3\simeq0.7$\\\hline\hline
      \end{tabular}
\end{center}
\end{table}
}

{\small
\begin{table}[hbt]
\begin{center}
\vspace*{-0.5cm} \caption{ \label{tab:radm1}\small $M1$ transitions
widths. $E_\gamma$ in MeV, $\Gamma$ in keV,
 Estimates of decay widths
containing mixing angles are given in terms of the same values used
in $E1$ transitions.}
 \vspace*{0.2cm}
\begin{tabular}{lcc}\hline\hline

Mode &$E_\gamma$& $\Gamma$\\\hline

$D(2550)^0[2\,^1S_0]\rightarrow D^{\ast 0}\gamma$ &477 & $13.9$\\

$D(2600)^0[2S-1D]\rightarrow D^0\gamma$ &638& $97.3c^2=82.6$\\

$D_{s1}(2710)[2S-1D]\rightarrow D_s\gamma$ &640 & $0.47c^2_2=0.03$\\
$D_{sJ}(2860)[2S-1D]\rightarrow D_s\gamma$ &754 & $0.88s^2_2=0.82$\\

$D_{sJ}(3040)[D_{s1}(2P)]\rightarrow D_{s1}(2460)$ &528
&$0.01c^2_3+0.08c_3s_3+0.1s^2_3=0.07$\\
$D_{sJ}(3040)[D^\prime_{s1}(2P)]\rightarrow D_{s1}(2460)$ &528
&$0.01s^2_3-0.08c_3s_3+0.1c^2_3=0.04$\\

$D_{sJ}(3040)[D_{s1}(2P)]\rightarrow D_{s1}(2536)$ &466
&$0.03c^2_3-0.04c_3s_3+0.01s^2_3=0.006$\\
$D_{sJ}(3040)[D^\prime_{s1}(2P)]\rightarrow D_{s1}(2536)$ &466
&$0.03s^2_3+0.04c_3s_3+0.01c^2_3=0.03$\\

$D_{sJ}(3040)[D_{s1}(2P)]\rightarrow D_{s0}(2317)$ &640
&$0.09c^2_3=0.06$\\
$D_{sJ}(3040)[D^\prime_{s1}(2P)]\rightarrow D_{s1}(2317)$ &640
&$0.09s^2_3=0.03$\\

$D_{sJ}(3040)[D_{s1}(2P)]\rightarrow D_{s2}(2573)$ &435
&$0.01c^2_3=0.007$\\
$D_{sJ}(3040)[D^\prime_{s1}(2P)]\rightarrow D_{s2}(2573)$ &435
&$0.01s^2_3=0.003$\\\hline\hline
      \end{tabular}
\end{center}
\end{table}
}

\section*{V. Summary and conclusion}
\indent \vspace*{-1cm}

  The discovery of $D(2550)$, $D(2600)$, $D(2750)$, $D(2760)$,
  $D_{s1}(2710)$, $D_{sJ}(2860)$, and $D_{sJ}(3040)$
   provides a good
opportunity to test our present understanding of charmed mesons and
is also of importance to further establish the $D$ and $D_s$
spectra. We are trying to shed some light on the natures of these
open-charm states by investigating their masses and decays in the
nonrelativistic constituent quark model.

We first calculated the charmed meson spectrum in the
nonrelativistic constituent quark model. The overall agreement
between our predicted masses and those from other approaches such as
the Blankenbecler-Sugar equation and the relativistic quark models
turns out to be satisfactory. According to the observed decay modes
and by comparing the measured masses with our predictions, we
presented the possible assignments of these newly observed
open-charm states.

Mass spectra alone are insufficient to determine the quantum numbers
of these open-charm states. Studies on the decay dynamics of these
states are needed. We therefore further evaluated the strong and
$E1$ and $M1$ radiative decays of these open-charm states for
possible assignments.

Comparing our predictions with the experiment, we conclude that (1)
if the $D(2550)$ is indeed the $D(2\,^1S_0)$ state, its width could
be overestimated experimentally; (2) the $D(2600)$ and
$D_{s1}(2710)$ can be identified as the $2\,^3S_1$-$1\,^3D_1$
mixtures; (3) if the $D(2750)$ and $D(2760)$ are the same state,
they could be interpreted as the $D(1\,^3D_3)$; otherwise, they
could be assigned as the $D(1\,^3D_3)$ and $D^\prime_2(1D)$,
respectively; (4) both the $D_s(1\,^3D_3)$ and $D_{s1}(2710)$'s
partner assignments for the $D_{sJ}(2860)$ are likely; and (5) both
the $D_{s1}(2P)$ and $D^\prime_{s1}(2P)$ interpretations for the
$D_{sJ}(3040)$ seem possible. Further experimental studies on these
states are needed.

 \section*{Acknowledgments}
 \indent\vspace{-1cm}

  We acknowledge Prof. Xiang Liu  for very helpful suggestions and discussions. This work
is supported in part by HANCET under Contract No. 2006HANCET-02, and
by the Program for Youthful Teachers in University of Henan
Province.

\end{document}